\documentclass{emulateapj}
\usepackage{apjfonts}
\DeclareGraphicsExtensions{.pdf,.png,.jpg}

\begin{document}

\title{PREDICTING THE YIELDS OF PHOTOMETRIC SURVEYS FOR TRANSITING EXTRASOLAR PLANETS}

\author{\sc Thomas G. Beatty\altaffilmark{1,2}
	 and B. Scott Gaudi\altaffilmark{3}
}

\altaffiltext{1}{Department of Physics and MIT Kavli Institute, Massachusetts Institute of Technology, 77 Massachusetts Ave., Cambridge, MA 02139}

\altaffiltext{2}{tbeatty@space.mit.edu}

\altaffiltext{3}{Department of Astronomy, The Ohio State University, 140 W. 18th Ave., Columbus, OH 43210}

\slugcomment{Accepted version for publication in the Astrophysical Journal}
\shorttitle{Predicting the Yields of Photometric Surveys}
\shortauthors{Beatty \& Gaudi}

\begin{abstract}
We develop a method for predicting the yield of transiting planets from a
photometric survey given the parameters of the survey (nights observed,
bandpass, exposure time, telescope aperture, locations of the target
fields, observational conditions, and detector characteristics), as well
as the underlying planet properties (frequency, period and radius
distributions).  Using our updated understanding of transit surveys
provided by the experiences of the survey teams, we account for those
factors that have proven to have the greatest effect on the survey yields.
Specifically, we include the effects of the surveys' window functions,
adopt revised estimates of the giant planet frequency, account for the
number and distribution of main-sequence stars in the survey fields, and
include the effects of Galactic structure and interstellar extinction.
We approximate the detectability of a planetary transit using a
signal-to-noise ratio (S/N) formulation.  We argue that our choice of
detection criterion is the most uncertain input to our predictions, and
has the largest effect on the resulting planet yield.  Thus drawing robust
inferences about the frequency of planets from transit surveys will
require that the survey teams impose and report objective, systematic, and
quantifiable detection criteria.  Nevertheless, with reasonable choices
for the minimum S/N, we calculate yields that are generally lower, more
accurate, and more realistic than previous predictions. As examples, we
apply our method to the Trans-Atlantic Exoplanet Survey, the XO survey,
and the {\it Kepler} mission. We discuss red noise and its possible
effects on planetary detections.  We conclude with estimates of the
expected detection rates for future wide-angle synoptic surveys.

\end{abstract}


\section{Introduction}
There are four ways by which extrasolar planets have been
detected. The first method to unambiguously detect an extrasolar
planet was pulsar timing, which relies on detecting periodic
variations in the timing of the received radio signal that occur as
the pulsar orbits about the system's barycenter. The first system of
three planets was found around PSR B1257+12 in 1992
\citep{wolszczan1992}, followed by a single planet around PSR B1620-26
\citep{backer1993}. Although rare, the pulsar planets are some of the
lowest mass extrasolar planets known: PSR B1257+12a is about twice the
mass of the Moon.

The second way to find extrasolar planets is through radial velocities
(RV), which uses the Doppler shift of observed stellar spectra to look
for periodic variations in the target star's radial velocity. By
estimating the mass of primary star, the observed radial velocity
curve and velocity semi-amplitude can then be used to directly
calculate the inclination-dependent mass ($M_p\sin i$) of the
companion object. To date, RV surveys have detected more than 230
planets around other stars, making it the most successful method of
extrasolar planet detection. While the large number of detected
systems having an unseen companion with a mass on the order of $1
M_{Jup} \sin i$ statistically ensures that the majority of these are
planetary bodies, the RV surveys are unable to provide more
information than the minimum masses, periods, eccentricities, and the
semi-major axes of the planets.

RV surveys also have a limited ability to detect planets much smaller
than a few Earth masses. The state of the art in RV surveys is the
High Accuracy Radial Velocity Planet Searcher (HARPS) spectrometer at
the La Silla Observatory in Chile, which is capable of radial velocity
measurements with precisions better than $1 \ \mathrm{ms^{-1}}$ for
extended periods of time \citep{lovis2006}. HARPS is therefore able to
detect planets with masses on the order of $3$ to $4 M_{\oplus}$ in
relatively short period orbits. Unfortunately, planets closer to an
Earth mass will be increasingly difficult to detect since intrinsic
stellar variability, in the form of acoustic oscillation modes and
granulations on the photosphere, makes more precise spectroscopic
radial velocity measurements harder to acquire. However, it may be
possible to surmount this obstacle (as in the case of
\citealt{lovis2006}) through the selection of stars with ``quiet''
photospheres and long integration times which serve to average out the
stellar variability.

Gravitational microlensing is another technique for detecting
extrasolar planets. Microlensing of a star occurs when a a star passes
near the line of sight of the observer to another background star. The
gravity of the foreground star acts as a lens on the light emitted by
the background star, which causes the star in the background to become
momentarily brighter as more light is directed towards the
observer. Planetary companions to the lens star can further magnify
the background star, and create short-term perturbations to the
microlensing light curve. To date, six planets have been detected
using microlensing \citep{bond2004, udalski2005, beaulieu2006,
gould2006a, gaudi2008}. Unfortunately, the one-shot nature of
microlensing observations means that information about systems
discovered this way is generally sparser than that available for RV
systems. Therefore, microlensing is most useful in determining the
general statistical properties of extrasolar planets, such as their
frequency and distribution, and not the detailed properties of the
planetary systems.

Planetary transits are a fourth method by which extrasolar planets
have been discovered, and the one that provides the most complete set
of information about the planetary system. Only planets with very
specific orbital characteristics have a transit visible from Earth,
because the orbital plane has to be aligned to within a few degrees of
the line of sight. Therefore transiting planets are
rare. Nevertheless, a transiting extrasolar planet offers the
opportunity to determine the mass of the planet (when combined with RV
measurements) since the inclination is now measurable, as well as the
planetary radius, the density, the composition of the planetary
atmosphere, the thermal emission from the planet, and many other
properties (see \cite{charbonneau2007} for a review). Additionally,
and unlike RV surveys, transiting planets should be readily detectable
down to $1 R_\oplus$ and beyond, even for relatively long periods.

Having accurate predictions of the number of detectable transiting
planets is immediately important for the evaluation and design of
current and future transit surveys. For the current surveys,
predictions allow the operators to judge how efficient are their
data-reduction and transit detection algorithms. Future surveys can
use the general prediction method that we describe here to optimize
their observing set-ups and strategies. More generally, such
predictions allow us to test different statistical models of
extrasolar planet distributions. Specifically, as more transiting
planets are discovered and characterized, predictions relying on
incorrect statistics of extrasolar planet properties will increasingly
diverge from the observed set of distributions.

Using straightforward estimates it appears that observing a planetary
transit should not be too difficult, presuming that one observes a
sufficient number of stars with the requisite precision during a given
photometric survey. Specifically, if we assume that the probability of
a short-period giant planet (as an example) transiting the disk of its
parent star is 10\%, and take the results of RV surveys which indicate
the frequency of such planets is about 1\% \citep{cumming08},
together with the assumption that typical transit depths are also
about 1\%, the number of detections should be $\approx 10^{-3}N_{\leq
1\%}$, where $N_{\leq 1\%}$ is the number of surveyed stars with a
photometric precision better than 1\%.

Unfortunately, this simple and appealing calculation fails. Using this
estimate, we would expect that the TrES survey, which has examined
approximately 35,000 stars with better than 1\% precision, to have
discovered 35 transiting short period planets. But, at the date of
this writing, they have found four. Indeed, overall only 51 transiting
planets have been found at this time by photometric surveys
specifically designed to find planets around bright stars\footnote{As
of June 2008. See the Extrasolar Planets Encyclopedia at
http://exoplanet.eu for an up-to-date list.}. This is almost one
hundred times less than what was originally predicted by somewhat more
sophisticated estimates \citep{horne2003}.

Clearly then, there is something amiss with this method of estimating
transiting planet detections. Several other authors have developed
more complex models to predict the expected yields of transit
surveys. \cite{pepper2003} examined the potential of all-sky surveys,
which was expanded upon and generalized for photometric searches in
clusters \cite{pepper2005}. \cite{gould2006b} and \cite{fressin2007}
tested whether the OGLE planet detections are statistically consistent
with radial velocity planet distributions. \cite{brown2003} was the
first to make published estimates of the rate of false positives in
transit surveys, and \cite{gillon2005} model transit detections to
estimate and compare the potential of several ground- and space-based
surveys.

As has been recognized by these and other authors, there are four
primary reasons why the simple way outlined above of estimating
surveys yields fails.

First, the frequency of planets in close orbits about their parent
stars (the planets most likely to show transits) is likely lower than
RV surveys would indicate. Recent examinations of the results from the
OGLE-III field by \cite{gould2006b} and \cite{fressin2007} indicate
that the frequency of short-period Jovian-worlds is on the order of
$0.45\%$, not $1.2\%$ as is often assumed by extrapolating from RV
surveys \citep{marcy2005a}. \cite{gould2006b} point out that most
spectroscopic planet searches are usually magnitude limited, which
biases the surveys toward more metal-rich stars, which are brighter
at fixed color. These high metallicity stars are expected
to have more planets than solar-metallicity stars \citep{santos2004,
fischer2005}.

Second, a substantial fraction of the stars within a survey field that
show better than 1\% photometric precision are either giants or early
main-sequence stars that are too large to enable detectable transit
dips from a Jupiter-sized planet \citep{gould2003, brown2003}.

Third, robust transit detections usually require more than one transit
in the data. This fact, coupled with the small fraction of the orbit a
planet actually spends in transit, and the typical observing losses at
single-site locations due to factors such as weather, create low
window probabilities for the typical transit survey in the majority of
orbital period ranges of interest \citep{vonbraun2007}.

Lastly, requiring better than 1\% photometric precision in the data is
not a sufficient condition for the successful detection of transits:
identifiable transits need to surpass some kind of a detection
threshold, such as a signal-to-noise ratio (S/N) threshold. The S/N of
the transit signal depends on several factors in addition to the
photometric precision of the data, such as the depth of the transit
and the number of data points taken during the transit
event. Additionally, ground-based photometry typically exhibits
substantial auto-correlation in the time series data points, on the
timescales of the transits themselves. This additional red noise,
which can come from a number of environmental and instrumental
sources, substantially reduces the statistical power of the data
\citep{pont2006}.

In this paper we describe a method to statistically calculate the
number of planets that a given transiting planet survey should be able
to detect. We have done so in a way that maximizes the flexibility of
our approach, so that it may be applied in as many circumstances as
possible. Therefore, we have allowed for survey-specific quantities
such as the observation bandpass to be defined arbitrarily, and have
also allowed for our astrophysical assumptions to be easily
altered. We account for factors such as the low frequency of Jovian
planets, the increased difficulty of detecting transits at high impact
parameters, as well as variations in stellar density due to Galactic
structure and extinction due to interstellar dust. We also use the
present-day mass function to realistically account for the number of
main-sequence stars that will be in a given field of view. We do not
treat, or simulate, the distribution and effects of giant stars. We
approximate the detectability of a planetary transit using a S/N
formulation.

This differs from the previously published work in the field by virtue
of the generalized nature of our approach. Other authors, such as
\cite{pepper2003}, \cite{gillon2005}, \cite{gould2006b}, and
\cite{fressin2007} have restricted themselves to modeling specific
transit surveys or types of transit surveys. Our approach, which
allows for variations such as arbitrary telescope parameters,
observing bandpasses, cadences, and fields of view, is not restricted
to any one type of transit survey. Indeed, we offer examples of how
our method may be applied to the four most frequently used modes for
transit searches. Furthermore, we have conducted an extensive
literature review of basic astrophysical relations, such as the
Present Day Mass-Function (PDMF), Galactic structure, and the main
sequence mass-luminosity relation, and included these in our
predictions. We do not use more complicated (but presumably more
accurate) Galactic models like the Besancon model \citep{robin1986}
because these models are less flexible and considerably more time
consuming to utilize.

We first describe the mathematical formalism with which we have chosen
to address this problem, and then move on to discuss our specific
assumptions. As examples, we then offer our predictions for several
different types of photometric surveys: point-and-stare,
drift-scanning, space-based, and all-sky. For each, we choose a
representative survey for that particular mode, and compare our
predictions to actual survey results, or to predictions elsewhere in
the literature. We also examine the possible effects of red noise on
our predictions, and conclude by discussing what our predictions imply
for the general state of photometric transiting planet surveys.

\section{General Formalism}
In the most general sense, we can describe the average number of
planets that a transit survey should detect as the probability of
detecting a transit multiplied by the local stellar mass function,
integrated over mass, distance, and the size of the observed field
(described by the Galactic coordinates $(l,b)$:
{\setlength\arraycolsep{0.1em}
\begin{eqnarray}\label{eq:10}
\frac{d^6 N_{det}}{dR_p\ dp\ dM\ dr\ dl\ db} &=& \rho_*(r,l,b)\ r^2 \cos b\ \frac{dn}{dM}\ \frac{df(R_p,p)}{dR_p\ dp} \nonumber \\ 
&&\times\ P_{det}(M,r,R_p,p),  
\end{eqnarray}
}
where $P_{det}(M,r,R_p,p)$ is the probability that a given star of
mass $M$ and distance $r$ orbited by a planet with radius $R_p$ and
period $p$ will present a detectable transit to the observing
set-up. $\frac{df(R_p,p)}{dR_p\ dp}$ is the probability that a star
will possess a planet of radius $R_p$ and period $p$. $dn/dM$ is the
present day mass function in the local solar neighborhood, and
$\rho_*$ is the local stellar density for the three-dimensional
position defined by ($r,l,b$). We use $r^2 \cos b$ instead of the
usual volume element for spherical coordinates, $r^2 \sin \phi$,
because $b$ is defined opposite to $\phi$: $b=90^\circ$ occurs at the
pole.

First, we will integrate over both planetary radius and orbital period
{\setlength\arraycolsep{0.1em}
\begin{eqnarray}\label{eq:15}
\frac{d^4 N_{det}}{dM\ dr\ dl\ db} &=& \rho_*(r,l,b)\ r^2 \cos b\ \frac{dn}{dM}\nonumber \\ 
&\times&\ \int_{R_{p,min}}^{R_{p,max}} \int_{p_{min}}^{p_{max}} \frac{df(R_p,p)}{dR_p\ dp}\nonumber \\
&\times&\ P_{det}(M,r,R_p,p) \,dR_p \,dp.
\end{eqnarray}
}
To do this, we must first specify the detection probability $P_{det}$,
as well as the distribution of planet radius and periods
$df(R_p,p)/dR_p d_p$, which we treat in Section 2.2.

\subsection{The Detection Probability $P_{det}$}
The probability that a planet orbiting a given star shows a detectable
transit may be broken up into two separate probabilities: the chance
that a planet with radius $R_p$ and period $p$ around a star of mass
$M$ at a distance $r$ will transit with a sufficient signal-to-noise
ratio to be detected, and the window probability that a transit will
be visible for the particular observing set-up:
\begin{equation}\label{eq:22}
P_{det}(M, r, R_p, p) = P_{S/N}(M, r, R_p, p)\ P_{win}(p)
\end{equation}

\subsubsection{$P_{S/N}$}

To calculate the probability that a planetary system will show
detectable transits, we first select the appropriate statistical test
to determine whether or not our notional data shows a discernible
transit. In the world of transiting planet searches, many different
methods are used either individually or together to test the
statistical significance of a possible transit and alert the
researcher to its existence. For example, the Hungarian-made Automated
Telescope Network (HATNet) uses a combination of the Dip Significance
Parameter (DSP) and a Box-Least Squares (BLS) power spectrum (amongst
others) to test its photometric data. In general though, whatever test
one uses, it is generally the S/N of a transit shape in the observed
data is what determines whether or not that transit will be identified
as an event.

As the square of the signal-to-noise can be approximated by the
$\chi^2$ test, we therefore chose to use the $\chi^2$ value to
determine if a given transit will be detectable for a set of given
observation parameters. For a specific transit, the $\chi^2$
statistic, assuming uncorrelated white noise,\footnote{We consider red
noise later, in Section 4.5.} is given by
\begin{equation}\label{eq:30}
\chi^2 = N_{tr} \left(\frac{\delta}{\sigma}\right)^2.
\end{equation}
Where $N_{tr}$ is the number of data points observed in a transit of
fractional depth $\delta$ and with fractional standard deviation
$\sigma$, assuming a boxcar transit shape with no ingress or
egress. To be detected, the transit must have a $\chi^2$ higher than a
certain minimum $\chi_{min}^2$
\begin{equation}\label{eq:40}
\chi^2 \geq \chi_{min}^2.
\end{equation}
The fractional depth of the transit is given by the square of the planet-to-star radius ratio 
\begin{equation}\label{eq:50}
\delta = \left(\frac{R_p}{R_*}\right)^2.
\end{equation}
While this formulation ignores the contributions of stellar
limb-darkening to the depth of the transit, \cite{gould2006b}
demonstrate that, to a first order approximation, transit detection
rates are not affected by limb-darkening. Briefly, they point out that
the effects of limb-darkening on the transit lightcurve will be offset
by the effects of the planet's ingress and egress from the stellar
disk. Therefore, we have not included stellar limb-darkening terms in
our calculations.

To determine $\sigma$, we assume Poisson statistics, such that for a
given exposure with an observed number of $N_S$ source photons from
the target star and $N_B$ background photons,
\begin{equation}\label{eq:60}
\sigma = \sqrt{\frac{N_S + N_B}{N_S^2} + \sigma_{scint}^2},
\end{equation}
where $\sigma_{scint}$ accounts for scintillation as per \cite{young1967}, and $N_S$ is defined as
\begin{equation}\label{eq:70}
N_S = e_{\lambda} F_{S,\lambda} t_{exp} A.
\end{equation}
Here $e_{\lambda}$ is the efficiency of the overall observing set-up,
and can take values from 0 to 1. It accounts for photon losses in
places such as the filter, the mirrors of the telescope, the CCDs, and
passage through the atmosphere. $F_{S,\lambda}$ is the source photon
flux in the observed bandpass in units of $\gamma\ m^{-2}\
s^{-1}$. $t_{exp}$ is the exposure time of each observation, and $A$
is simply the light collecting area of the telescope.

$F_{S,\lambda}$, the observed photon flux from the source in a certain
bandpass, can be defined as a function of the photon luminosity of the
source, the distance to the source, and the effects of extinction due
to interstellar dust:
\begin{equation}\label{eq:80}
F_{S,\lambda} = \frac{\Phi_{\lambda}(M)}{4\pi r^2} 10^{-A_\lambda / 2.5},
\end{equation}
where $10^{-A_\lambda / 2.5}$ accounts for extinction, and will be
treated later. It has the effect of decreasing the number of observed
photons from more distant stars, particularly in the bluer bands.

$\Phi_{\lambda}(M)$, the photon luminosity of the source star, we set
as a function of the stellar mass $M$ and the observational bandpass
$\lambda$. We derive this quantity in Appendix A, and only note here
that we approximate the source stars as blackbodies. While this is a
simplification, we show in Appendix A that it is sufficient for our
purposes. Using $\Phi_{\lambda}(M)$ in equation (\ref{eq:80}) yields
the observed photon flux at a distance $r$ from that same star.

We thus now have an expression for $N_S$ as a function of the stellar
mass, distance, and bandpass choice, that we can use in our
calculation of $\sigma$. We now only need $N_B$, the number of
background photons observed. Similar to equation (\ref{eq:70}), $N_B$
is defined as
\begin{equation}\label{eq:140}
N_B = S_{sky,\lambda} \Omega t_{exp} A,
\end{equation}
where $S_{sky,\lambda}$ is the photon surface brightness of the sky at
the particular wavelength $\lambda$ (or, in our case, for a specific
band), and $\Omega$ is the effective area of the seeing disk. Assuming
that the point-spread function (PSF) is a Gaussian with a full-width
half-max (FWHM) of $\Theta_{FWHM}$ arcseconds, $\Omega$ is defined by
\begin{equation}\label{eq:150}
\Omega = \frac{\pi}{\ln 4} \Theta_{FWHM}^2.
\end{equation}
We can now place both of the expressions for $N_S$ and $N_B$ back into equation (\ref{eq:60}) to arrive at a determination of $\sigma$, the photometric uncertainty in the data points.

In certain cases, $N_S$ and $N_B$ will be so great that the CCD
detectors of the telescope will be saturated by the number of photons
received from a star during the exposure time. In this case, further
measurements become useless. Mathematically, we may describe
saturation as occurring when the number of photons collected by a
single CCD pixel during an exposure exceeds the Full-Well Depth (FWD)
of that pixel:
\begin{equation}\label{eq:75}
N_{pixel} \geq N_{FWD}.
\end{equation}
$N_{pixel}$ is related to $N_S$ and $N_B$ as
{\setlength\arraycolsep{0.1em}
\begin{eqnarray}\label{eq:76}
N_{pixel} &=& N_S \left(1-\mathrm{exp}\left[-\ln 2 \left(\frac{\Theta_{pix}}{\Theta_{FWHM}}\right)^2\right]\right)\nonumber \\
 &&+ \frac{N_B \Theta_{pix}^2}{\Omega}, 
\end{eqnarray}
}
where $\Theta_{pix}$ is the angular size of an individual pixel.

The only part of equation (\ref{eq:30}) that remains to be defined is
the number of data points observed while the system is in transit,
$N_{tr}$. Assuming a circular orbit, we can say that the number of
points seen in transit is directly proportional to the fraction of
time that the system spends in eclipse, and given a total number of
observed data points $N_{tot}$, $N_{tr}$ is thus
\begin{equation}\label{eq:160}
N_{tr} = N_{tot} \left(\frac{R_*}{\pi a} \right)\sqrt{1-b^2},
\end{equation} 
where $b$ is the impact parameter of the transit in units of the
stellar radius, $b\equiv(a/R_*)\cos i$, $i$ is the orbital
inclination, and we have assumed that $R_*<<a$. We note that this does
not hold for correlated noise, see \cite{aigrain2007}.

Taking equation \ref{eq:160} and putting it back into our original
formulation for $\chi^2$ gives
\begin{equation}\label{eq:170}
N_{tot} \left(\frac{R_*}{\pi a} \right)\sqrt{1-b^2} \left(\frac{\delta}{\sigma}\right)^2 \geq \chi_{min}^2. 
\end{equation} 
We define a new variable, $\chi_{eq}^2$, as the value of equation
(\ref{eq:170}) when $b=0$, corresponding to the $\chi^2$ of an
equatorial transit about a given star. This gives us a new form that
makes it easier to determine when a system will have a transit
geometry sufficient to be detected by the given observing set-up:
\begin{equation}\label{eq:180}
\chi_{eq}^2 \sqrt{1-b^2} \geq \chi_{min}^2. 
\end{equation} 
Rearranging, we find that we are able to detect transits up to a $b_{max}$ of
\begin{equation}\label{eq:190}
b_{max}(M, r, p) = \sqrt{1-\left(\frac{\chi_{min}^2}{\chi_{eq}^2}\right)^2}.
\end{equation} 

Therefore, $b_{max}$ is a function of $M$, $r$, and $p$ as a result of
the relations contained within the $\chi_{eq}^2$ term.

Finally, we can now deduce an explicit expression for $P_{S/N}$. Given
that detectable transits can only occur between $0 \leq b \leq
b_{max}$, we may simply integrate the normalized probability density
function of allowable impact parameters from 0 to $b_{max}$ to find a
value for $P_{S/N}$.

Given that $b\equiv(a/R_*) \cos i$, the maximum value of the
impact parameter for a circular orbit (different from the maximum at
which we will detect transits) is $b=a/R_*$.  The normalization for
the impact parameter's probability density function is therefore
\begin{equation}\label{eq:200}
1 = \int_0^{\frac{a}{R_*}} P_b(b) \,db.
\end{equation} 
Since impact parameters are uniformly distributed\footnote{The
distribution of $\cos i$ is uniform, which follows from the assumption
that planetary systems are randomly oriented. The distribution of
$\cos i$ may be derived by considering the normalized angular momentum
vector of the planetary orbit, which is free to point in any
direction, and thus has a parameter space that forms an imaginary
sphere around the parent star. Particular values of $\cos i$ are
``hoops'' along the surface of this sphere, and the ratio of their
area to that of the total sphere is the probability that $\cos i$ will
take that value.}, this gives the solution $P_b = R_*/a$. It
should be noted that this form of $P_b$ is only valid in the case of
circular orbits. For the period ranges probed so far by transit
surveys, this is not a very restrictive assumption, since only 
$\sim 10\%$ of the known transiting planets have eccentricities inconsistent with
zero. At longer periods, however, the effects of eccentricity must
be considered \citep{barnes07,burke08}.

If we now integrate this over the range of impact parameters that yield detectable transits, we find
{\setlength\arraycolsep{0.1em}
\begin{eqnarray}\label{eq:210}
P_{S/N}(M,r,R_p,p)&=&\int_0^{b_{max}} \frac{R_*}{a} \,db \nonumber \\
                      &=&\frac{R_*}{a} \sqrt{1-\left(\frac{\chi_{min}^2}{\chi_{eq}^2}\right)^2}.
\end{eqnarray}
}

\subsubsection{$P_{win}$}

The window probability (or window function) for a transit survey
describes the probability that a transiting planet with a certain
period will be visible to the survey, given the observational cadence
of the survey (i.e. - X hours of observing per night for Y
nights). While many numerical routines exist that can exactly
calculate the window probability for a given observing set-up, for our
purposes calculating an exact numerical answer is generally too time
intensive to be practical.

\begin{figure}
\vskip -0.0in 
\epsscale{1.2} 
\plotone{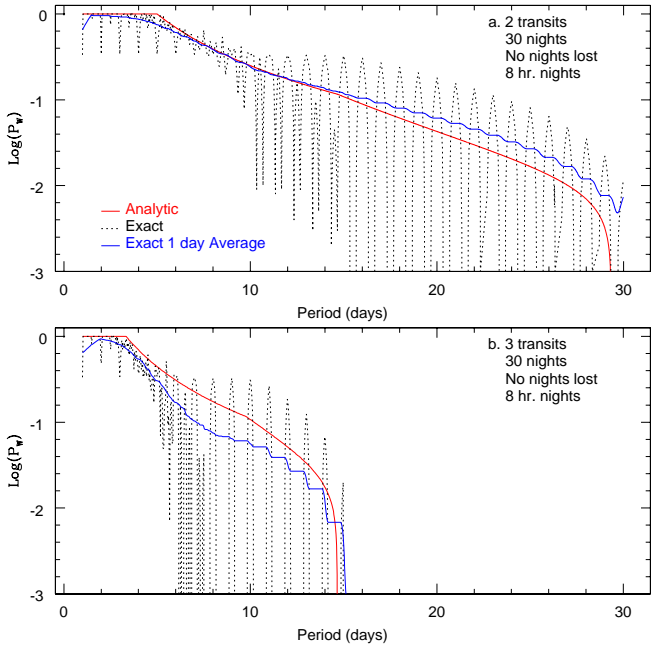}
\vskip -0.0in 
\figcaption[]{Comparing the analytic and exact formulations of the window probability. The top panel shows the window probability for detecting 2 transits over a 30 day observing run with no nights lost and 8 hours of observing per night. The bottom panel uses the same situation, but is instead for a minimum of three transits.}
\end{figure}

 We therefore developed an analytical framework that allows for the
approximate calculation of a window probability. This has the virtue
of requiring much less computing time. This is gained at the cost of a
small loss of accuracy.

We give a full description of our analytic window probability in
Appendix B. Briefly, we assume randomly sampled observations, which
leads to simple expressions for the average window probability. This
implicitly ignores the effects of aliasing at integer and half-integer
day periods. Thus we are not able to address issues related to
non-uniform or biased \emph{a posteriori} distributions of planet
parameters resulting from the effect of aliasing
(e.g. \cite{pont2006}). Nevertheless, the analytic solution does
follow the ``outline'' of the exact calculations, and closely tracks
the moving average value of the exact calculations sufficiently
accurately for our purposes, as is shown in Figure 1.

\subsection{The Frequency of Planets $\frac{df(R_p,p)}{dR_p\ dp}$}

$df(R_p,p)/(dR_p\ dp)$ denotes the probability that a given star
has a planet with radius $R_p$ and orbital period $p$.
In the example surveys that we consider later, we chose to focus on
the detection of short period Very Hot- and Hot-Jupiters (VHJs and
HJs) with orbital periods of 1 to 5 days, as this is the regime
being probed by these surveys.  The exception of this focus
is the Kepler survey, where we also examine the detectability of
transiting planets over a range of radii and orbital periods. In both
cases, we consider a fixed planet radius, so that
\begin{equation}\label{eq:230}
\frac{df(R_p,p)}{dR_p\ dp} = k(p) f(p)\delta(R_p - R_p^\prime),
\end{equation}
where $k(p)$ is the correct normalization for a given period range, $f(p)$
is the distribution within this range, and
$R_p^\prime$ is the planet radius. For VHJs and HJs we use
$R_p^\prime = 1.1 R_{Jup}$, the average radius of the transiting
planets found to date\footnote{http://www.inscience.ch/transits/}.
In general,
we have left the exact distribution of $f(p)$ to
be assigned as is deemed appropriate.  For the VHJ and HJ period distributions
described above, we adopt a distribution that is uniform in period, $f(p)=1$.

The frequency of Hot Jupiters ($P\le 5~{\rm days}$, $M_p=0.1-10\ M_{Jup}$) 
is commonly taken to be $\sim 1\%$ from the results of the RV planet surveys
(e.g., \citealt{marcy2005a}).   However, adopting frequencies from
RV surveys can lead to biased
estimates of the expected yields from transit surveys, 
because RV surveys constrain the frequency of planets as a function of
mass, rather than radius.  Furthermore, as argued by \citet{gould2006b}, 
most RV surveys are more metallicity biased than transit surveys, and therefore
the planet frequency found by RV surveys will be correspondingly higher, 
since planet occurrence is an increasing function of metallicity \citep{santos2004,fischer2005}.
RV surveys are metallicity biased 
because the target samples for these surveys are magnitude-limited.  Since
metal rich stars are brighter at fixed color, they are overrepresented in
magnitude-limited samples with respect to a volume-limited sample \citep{fischer2005}.
Indeed, \cite{marcy2005b} note that because they use a
magnitude-limited target list, their target stars have a definite
metallicity bias.
Transit surveys, on the other hand, are typically signal-to-noise ratio limited.  At fixed
color, metal-rich stars are both brighter and bigger.  These two effects on
the signal-to-noise ratio roughly cancel, and as a result transit surveys are very weakly biased
with respect to metallicity (see Section 8.1
of \citealt{gould2006b} for a quantitative discussion.)

Indeed, there is evidence for the effects of this metallicity bias in
the comparison of various surveys. From the California and Carnegie
Keck magnitude-limited and metallicity-biased \citep{fischer2005,marcy2005b} survey sample,
\citet{cumming08} find a frequency of $\sim 1.3\%$ for planets with $P\le
5~{\rm days}$ and $M\ge 0.2 M_{Jup}$.  On the other hand, for the
volume-limited \citep{udry2000b} CORALIE sample, \citet{udry2007}
report a frequency of $\sim 0.8\%$ for $M\ge 0.2 M_{Jup}$ and $P\la
11.5~{\rm days}$ $(a< 0.1~{\rm AU})$.  Assuming that $\sim 75\%$ of
these planets have $P\le 5~{\rm days}$, this corresponds to a frequency
of 0.6\% for $P\le 5~{\rm days}$ and $M\ge 0.2 M_{Jup}$, a factor of
$\sim 2$ lower than that found for the Keck RV survey.
\citet{gould2006b} examined the OGLE-III transit survey, and found a
frequency of $\sim 0.5\%$ for $P\le 5~{\rm days}$, which was
subsequently confirmed and strengthened by \citep{fressin2007}.  Thus, as would
be expected based on the arguments above, the frequencies inferred from transit surveys
and volume-limited RV surveys are similar (and entirely consistent considering
Poisson uncertainties and uncertainties in the mass-radius relation of 
giant planets), whereas they are substantially lower than the
frequencies inferred from magnitude-limited (and metallicity-biased) RV
surveys. 
 
For the VHJ and HJ predictions that we later make using TrES, XO, and
Kepler, we will adopt the normalizations from \citet{gould2006b} of
$k(p) = 1/690$ for VHJs with orbital periods uniformly distributed
between 1-3 days, and $k(p) = 1/310$ for HJs uniformly distributed
between 3-5 day periods.  By using statistics from OGLE-III, instead
of the RV statistics, we are avoiding the difficulties discussed
above, and furthermore any unaccounted-for biases inherent in
photometric surveys will be preserved, and therefore any systematic
effects unknown to us will still be accurately captured in the final
predictions. In other words, it is an ``apples-to-apples'' comparison,
instead of ``apples-to-oranges'' using RV data.

\subsection{The Present Day Mass Function}

The next step is to integrate equation (\ref{eq:10}) over mass:
\begin{equation}\label{eq:240}
\frac{d^3N_{det}}{dr\ dl\ db} = \rho_*(r,l,b) r^2 \cos b \int_{M_{min}}^{M_{max}} P_{det}(M,r) \frac{dn}{dM} \,dM
\end{equation}

Having previously derived an expression for $P_{det}$, we need only determine the appropriate formulation for the PDMF, $\frac{dn}{dM}$. Using the results of \cite{reid2002}, who used the Hipparcos data sets along with their own data from the Palomar/Michigan State University survey to create a volume-limited survey out to 25 pc, we adopt
\begin{eqnarray}\label{eq:250}
\frac{dn}{dM} &=& k_{norm} \left(\frac{M_*}{M_\odot}\right)^{-1.35}\ \mbox{ for } 0.1 \leq M_*/M_\odot \leq 1,\\
\frac{dn}{dM} &=& k_{norm} \left(\frac{M_*}{M_\odot}\right)^{-5.2}\ \ \mbox{ for } 1 < M_*/M_\odot, \nonumber
\end{eqnarray}
where $k_{norm}$ is the appropriate normalization constant. A fuller discussion for our decision to use this particular form for the PDMF is in Appendix C. 

Adopting a stellar mass density in the Solar Neighborhood of $0.032 M_\odot \ \mathrm{pc}^{-3}$ \citep{reid2002}, we find a normalization of $k_{norm} = 0.02124\ \mathrm{pc}^{-3}$.

\subsection{Galactic Structure} 

While our normalization of the mass function properly describes the density of stars in the Solar Neighborhood, beyond about 100 pc we must account for variations in stellar density due to Galactic structure. Additionally, as mentioned previously, we must also take into consideration the effects of extinction due to interstellar dust. This is the final stage of integration that needs to be done on equation (\ref{eq:10}):
\begin{equation}\label{eq:260}
N_{det} = \int_{0}^{r_{max}} \int_{l_{min}}^{l_{max}} \int_{b_{min}}^{b_{max}} P_{det}(r) \rho_*(r,l,b) r^2 \cos b \,dr \,dl \,db.
\end{equation}

\subsubsection{Galactic Density Model}

We adopt the Bahcall \& Soneira model for the Galactic thin disk \citep{bahcall1980}. The model treats stellar density in the thin disk as a double exponential function involving both the distance from the Galactic center and height above the Galactic plane. Specifically, the mass density relative to the local density is given by
\begin{equation}\label{eq:280}
\rho_*(r,l,b) = exp\left[-\frac{d - d_{gc}}{h_{d,*}} - \frac{|z|}{h_{z,*}}\right],
\end{equation}
where $d_{gc}=8$ kpc is the distance from the Galactic Center. $h_{d,*}$ is the scale length, and $h_{z,*}$ is the scale height.

We adopt a value of $h_{d,*} = 2.5\ \mathrm{kpc}$ for the scale length of the disk. For the vertical scale height, we use the following relation, which is a slightly modified form of that originally proposed by \cite{bahcall1980}:
\begin{eqnarray*}\label{eq:285}
h_{z,*} &=& 90\ \ \ \ \ \ \ \ \ \ \ \ \ \ \ \ \ \ \ \ \ \ \ \ \ \mbox{ for } M_V \leq 2,\\
        &=& 90 + 200 \left(\frac{M_V-2}{3}\right)\ \mbox{ for } 2 < M_V < 5, \nonumber \\
        &=& 290\ \ \ \ \ \ \ \ \ \ \ \ \ \ \ \ \ \ \ \ \ \ \ \ \mbox{ for } 5 \leq M_V. \nonumber
\end{eqnarray*}
We calculate the absolute V-band magnitudes of stars of a given mass using the blackbody assumptions, and the mass-radius and mass-luminosity relations described in Appendix A.

We defer the complete discussion of our reasoning behind our choices for $h_{d,*}$ and $h_{z,*}$ to Appendix D. The exact values for these parameters are not well constrained in the literature, and we show the effects of using different values in the Appendix. We note that within the range of values found in the literature, the final prediction results may differ by as much as 20\%.

We only treat the thin disk in our modeling because we decided that an expanded analysis that also looked at stars in the thick disk, halo, and bulge would not add a significant number of detections to our final result. Both the thick disk and the halo have much lower volume densities than the thin disk at distances typically surveyed by transit searches. In the same vein, the bulge is too far away. These populations are also typically older than the thin disk, and so have a fainter turn-off point, meaning that these stars are more difficult to detect. That being said, for deeper transit surveys, such as some of the all-sky surveys, these populations may play an interesting role. In that case, it would be fairly simple to add them into our model. 

We chose the Bahcall model instead of other more detailed models due to speed and portability considerations. While others, such as the Besancon models \citep{robin1986}, take into account features such as density variations from the spiral structure of the Galaxy and the Galactic Bulge, they typically require over 30 minutes to simulate a given star field. Additionally, many are only accessible through webpage interfaces, and therefore difficult to include in a self-contained prediction algorithm.

We do note that comparisons between our Galactic model and the Besancon model revealed star count differences of only 10\% in the Kepler and TrES fields over a range of magnitude limits.

\subsubsection{Interstellar Extinction Modeling}

We also model the density of interstellar dust as a double exponential
\begin{equation}\label{eq:290}
\rho_{dust}(r,l,b) = exp\left[-\frac{d - d_{gc}}{h_{d,dust}} - \frac{|z|}{h_{z,dust}}\right].
\end{equation}
Which gives the dust density at a distance $r$ along a particular line of sight ($l,b$) relative to the density in the Solar Neighborhood (i.e. - $z=0$, $d=d_{gc}$). To calculate the extinction in a given bandpass for a star at a given distance, we then integrate equation (\ref{eq:290}) over $r$ to arrive at the effective path length through an equivalent Solar Neighborhood dust density, 
\begin{equation}\label{eq:310}
\ell_{dust} = \int_0^r \rho_{dust}(r,l,b) \, dr.
\end{equation}
The V-band extinction in magnitudes, $A_V$, is given by this path length multiplied by the appropriate normalization constant: $A_V = k_{dust,v} \ell_{dust}$. We adopt $k_{dust,v} = 1$ mag kpc$^{-1}$ \citep{struve1847}. We determine the extinction in other bands by using the extinction ratios ($A_\lambda/A_V$) from \cite{cox2000}, assuming that $R_V = 3.1$. 

\section{Assumptions}

It is important to keep in mind that in the preceding derivations, we made several assumptions about the nature of extrasolar planetary systems, as well as some simplifications to keep the math and computing time more manageable.

\medskip
1. {\em Detectability is Best Measured Through $\chi^2$:} While S/N is a better metric than photometric precision to determine if a transit will be detectable, and the $\chi^2$ parameter is a good proxy for S/N, actual transit surveys use a wide variety of tests to detect transits. Many tests, such as the visual evaluation of the data by a human being, are difficult to simulate numerically. Thus, using just a $\chi^2$ test is a simplification, as human inspection may impose a much higher effective S/N threshold. As we show for typical surveys, the number of detections is a strong function of the S/N threshold. Thus a proper determination of the effective S/N threshold imposed by all cuts is necessary before any robust conclusions can be drawn about the rate of detections relative to expectations. This is a fundamental - and unavoidable - limitation of our study.

\medskip
2. {\em No Stellar Limb-Darkening:} We have neglected to include terms for stellar limb-darkening because, as stated above, to first-order its effects are canceled by the the effects of ingress and egress on the transit lightcurve \citep{gould2006b}.

\medskip
3. {\em Stars are Blackbodies:} We assume that the radiated energy of a star can be described as a blackbody. While in general this is a fair approximation (to within 0.5 mag, see Figures 11 and 12), it does not hold true for the bluer  bandpasses we consider (U-band in particular) and for very low-mass stars.

\medskip
4. {\em Simplified Galactic Structure:} While we do treat variations in stellar density and interstellar dust densities in a general sense, our formalism is a simplified double exponential model that does not account for the spiral structure of the Galaxy or patchiness in the dust distribution. Therefore detailed, accurate predictions for an individual field may be better served by incorporating more sophisticated Galactic models (e.g. \cite{fressin2007}).

\medskip
5. {\em No Binary Systems:} 
While we only consider main sequence stars, we do not model the
statistics of binary systems; in our formulation all stars are treated
as single. Though this is clearly not the case, it was outside the
scope of this project to go deeply into binary frequencies, the
ability of planets to form in such systems, and whether or not such
planets would be detectable. 

However, we can make a rough estimate of the expected magnitude of the effect of binaries.  First,
we assume that 1/3 of our stars have no binary companion \citep{duquennoy91}.
We assume the remaining 2/3 of the stars are in binary systems.  A typical wide-field survey has a PSF
size of $\sim 20''$, which corresponds to a separation of $\sim 4000~{\rm AU}$, and a period
of $\log (P/{\rm day})\sim 7.8$ for an equal-mass solar-mass binary.
Adopting the period distribution for binary stars from \citet{duquennoy91}, 
we estimate that $\sim 90\%$ of the binary systems are unresolved.  Of these unresolved
binaries, we conservatively assume that those with separations $\la 5~{\rm AU}$, or
$\log (P/{\rm day})\sim 3.5$, do not host hot Jupiters because the binary
components are too close.  

Thus we estimate that $\sim 62\%$ of 
unresolved binaries can host hot Jupiters.  To estimate the detectability of transiting
planets in these,
consider a binary where $F_1$, $R_1$, $M_1$ and $F_2$, $R_2$, $M_2$ are the flux, radius, and 
mass of the primary (``1'') and secondary (``2''), respectively.  Additionally, define $\ell=F_2/F_1$,  
${\cal R}=R_2/R_1$, and $q=M_2/M_1\le 1$.  Then, for fixed planet radius and semimajor axis, assuming
photon-limited uncorrelated uncertainties, the signal-to-noise
ratios if the stars are resolved are $({\rm S/N})_1 \propto R_1^{-3/2} F_1^{1/2}$
and $({\rm S/N})_2 \propto R_2^{-3/2} F_2^{1/2}$, while if the stars are unresolved the
signal-to-noise ratios for transits across each component are
$({\rm S/N})_{B,1} \propto R_1^{-3/2} F_1^{1/2} (1+\ell)^{-1/2}$ and $({\rm S/N})_{B,2} \propto R_2^{-3/2} F_2^{1/2} (1+\ell)^{-1/2}$.
Assuming a constant volume density of stars and no interstellar extinction, the number of detected
planets scales as $({\rm S/N})_{\rm min}^{-3}$ \citep{gould2003b}.  Thus the number of planet detections for a
population of binaries in the unblended case
can be written as $N_0 \propto ({\rm S/N})_1^3 + f(q) ({\rm S/N})_2^3$, where $f(q)$ is the mass ratio distribution.
Similarly, for the blended case, this is $N_B \propto ({\rm S/N})_{B,1}^3 + f(q) ({\rm S/N})_{B,2}^3$.  Using the
scalings for the signal-to-noise ratios  just derived,
we evaluate the ratio of the number of detections in the blended case, relative to the number of detections
one would expect if they were unblended,
\begin{equation}
\frac{N_B}{N_0} = (1+\ell)^{-3/2} \frac{1+f(q)\ell^3{\cal R}^{-9/2}}{1+f(q)\ell^{3/2}{\cal R}^{-9/2}}.
\label{eq:291}
\end{equation}
We adopt $\ell = q^5$, which is roughly appropriate for the $R$-band (see Fig. 11),
and ${\cal R} = q^{0.8}$ (Eq.\ \ref{eq:121}).  We consider two different distributions
for $f(q)$.  First, we adopt $f(q)$ from \citet{duquennoy91}, which is weighted toward
low mass ratio systems.  Second, we consider a distribution with a substantial 
population of `twins' with $q=1$.  Specifically, we choose $f(q)\propto 1 +3.5\exp[-0.5(q-1)^2/\sigma_q^2]$,
with $\sigma_q=0.2$, which approximates the distribution found by \citet{halbwachs03}.
When we then average Equation (\ref{eq:291}) over $q$,
weighting by $f(q)$,  we find $\langle N_B/N_0\rangle \simeq 0.89$ for the \citet{duquennoy91} mass ratio distribution, and 
$\langle N_B/N_0\rangle \simeq 0.68$ for the \citet{halbwachs03} mass ratio distribution.  
Thus, we estimate that roughly 70-90\% of planets in unresolved
binaries will still be detectable.  The basic point is that, because the mass-luminosity
relation is so steep, only those binaries that have $q\sim 1$ result in a substantial
reduction in the signal-to-noise ratio.  Low mass ratio companions
have little affect on the ability to detect transit signals. For the
mass ratio distribution of \citet{duquennoy91}, these provide the bulk
of the companion population, and so in this case the effects of binarity are suppressed
even further.

Finally, we can summarize these various components to
provide a rough estimate of the fraction of planets that will be detected when we consider
binary systems, relative to our fiducial assumption that all stars are single.  
We find $1/3 + (0.1+ 0.62 \langle N_B/N_0\rangle)\times (2/3) \simeq 0.68-0.77$.  Thus we expect that our 
estimated yields are systematically high by at most $30\%$. This is generally of order or smaller
than other sources of uncertainty in our predicted yields.   

We note that a corollary of our assertion that binaries have a small
effect on the yields of transit surveys is that a substantial fraction of the
planets that have been found in wide-field transit surveys should be
in binary systems.  Only two planets in confirmed common proper-motion
binary systems have been reported \citep{burke2007,bakos2007}, however
systematic surveys for stellar companions to the host stars of the
known transiting planets have not been performed, and therefore the
true incidence of binary systems amongst the sample of transiting
planets is not known.

\section{Results}

Using the preceding formalism, we simulated four different surveys modes used by transit searches: point-and-stare, drift-scanning, space-based, and all-sky surveys. For the point-and-stare and drift-scanning cases, we use the TrES and XO transit surveys as our examples, respectively. Both surveys have been successful in their hunt for planets, and so we are able to directly compare the results of our calculations with the actual number of detections. For VHJs and HJs, we simulate the detection of planets with $R_p=1.1R_{Jup}$, and with the period distribution from \cite{gould2006b} that we describe in Section 2.2.

\begin{deluxetable}{c|c}
\tablecaption {\sc TrES Survey Parameters}
\tablewidth{0pt}
\tabletypesize{\small}
\startdata
\hline
\hline
Band & 550-700 nm\\
$m_R$ Limit & $\approx 13$\\ 
Observation Time & Varies\\
Exposure Time & 90 seconds\\
CCD Read Time & 25 seconds\\
Telescope Diameter & 0.1 m\\
Throughput & 0.6\\
PSF (FWHM) & 25''\\
$\theta_{pix}$ & 10''\\
CCD Full-Well Depth & $1.2\times10^5\ e^-$\\
\hline
\hline
Field of View & $36\ \mathrm{deg}^2$\\
Galactic Longitude & Varies\\
Galactic Latitude & Varies
\enddata
\end{deluxetable}

\begin{deluxetable}{c|cc}
\tablecaption {\sc Properties of the Simulated TrES Survey Fields}
\tablewidth{0pt}
\tabletypesize{\small}
\startdata
\hline
\hline
Field & Observation Time & Galactic Coordinates\\
\hline
And0 & 765 hrs & [126.11, -015.52]\\ 
Cyg1 & 609 hrs & [084.49, +010.28]\\ 
Cas0 & 628 hrs & [120.88, -013.47]\\ 
Per1 & 643 hrs & [156.37, -014.04]\\ 
UMa0 & 497 hrs & [168.87, +047.70]\\ 
Crb0 & 537 hrs & [053.49, +048.92]\\ 
Lyr1 & 582 hrs & [077.15, +017.86]\\ 
And1 & 435 hrs & [109.03, -017.62]\\ 
And2 & 332 hrs & [115.52, -016.21]\\ 
Tau0 & 406 hrs & [169.83, -015.94]\\ 
UMa1 & 500 hrs & [156.32, +054.01]\\
Her1 & 592 hrs & [063.26, +026.42]\\
Lac0 & 686 hrs & [097.79, -008.86]
\enddata
\end{deluxetable}

We demonstrate our method for space-based surveys by simulating the upcoming Kepler mission, and comparing our predictions with other predictions for Kepler in the literature. Similarly for the all-sky mode, we examine the future planned observing programs of LSST, SDSS-II, and Pan-STARRS and make predictions for each.

For each mode, we have tried to approximate the effect of red noise on detectability by increasing our S/N cut \citep{pont2006}. To show the effects of a more accurate red noise calculation, in Section 4.5 we evaluate its effects on the TrES survey.

\subsection{Point-and-stare - TrES}

Point-and-stare observing is the traditional observing mode of
tracking one field of sky continuously for all, or part, of the
night. There are several operational photometric transit surveys that
use this particular method of observing, including HATnet
\citep{bakos2004}, SuperWasp \citep{pollacco2006}, and KELT
\citep{pepper2007}, but we chose to model is the Trans-Atlantic
Exoplanet Survey (TrES). TrES is composed of a network of three
telescopes: STARE on the Canary Islands, PSST at Lowell Observatory in
Arizona, and Sleuth on Mt. Palomar, California. All are small, 10 cm
aperture, wide-field ($6^\circ$), CCD cameras that operate in unison
to observe the same field of sky nearly continuously over one to two
month periods. The survey specifics are described in more detail by
\cite{dunham2004} and by \cite{brown1999}. The relevant quantities for
our modeling are listed in Table 1.

\begin{figure}
\epsscale{1.2} 
\plotone{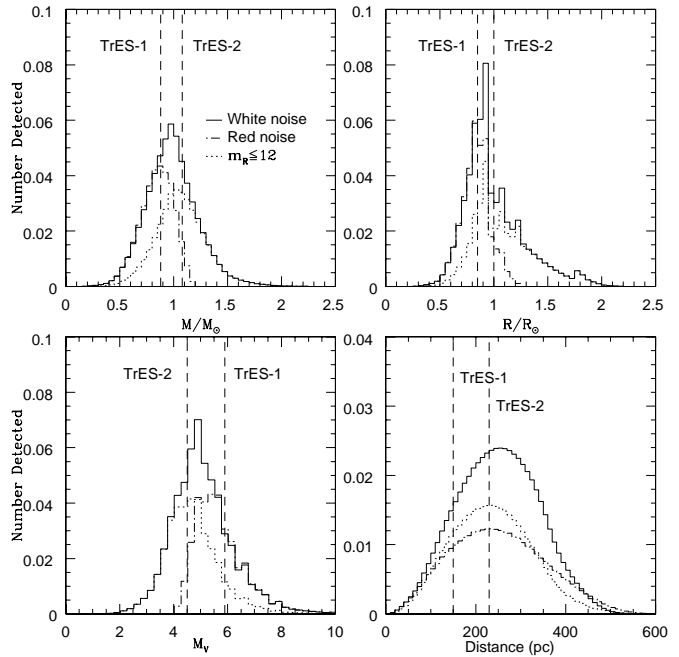}
\vskip -0.0in 
\figcaption[]{The distribution of predicted TrES detections in the Lyr1 field with and without red noise, and assuming a brighter magnitude limit than our fiducial value of $m_R\leq13$. The location of the stars TrES-1 and -2 are also shown. TrES-2 lies within Lyr1, while TrES-1 is in the nearby Lyr0 field.}
\end{figure}

We fit to the RMS plots posted on the Sleuth observing
website\footnote{http://www.astro.caltech.edu/~ftod/tres/sleuthObs.html}
to determine the background sky brightness in the TrES bandpass
(550-700nm), using the TrES team's stated photon throughput of 0.6,
and we estimate that the TrES bandpass has a sky background of $19.6\
\mathrm{mag\ arcsec}^{-2}$.

\begin{deluxetable*}{c|ccccccccccccccc|c}
\tablecaption {\sc Predicted TrES Detections for $m_R\leq13$, S/N$\geq30$, and $R_p=1.1R_{Jup}$ }
\tablewidth{0pt}
\tabletypesize{\scriptsize}
\startdata
\hline
\hline
     & And0 & Cyg1 & Cas0 & Per1 & UMa0 & CrB0 & Lyr1 & And1 & And2 & Tau0 & UMa1 & Her1 & Lac0 & Total\\
\hline
VHJs & 0.38 & 0.37 & 0.34 & 0.34 & 0.18 & 0.21 & 0.32 & 0.24 & 0.19 & 0.22 & 0.17 & 0.21 & 0.41 & 3.59\\
HJs  & 0.34 & 0.30 & 0.29 & 0.28 & 0.16 & 0.18 & 0.27 & 0.18 & 0.09 & 0.15 & 0.16 & 0.19 & 0.34 & 2.91\\
\hline
Both & 0.72 & 0.67 & 0.63 & 0.62 & 0.34 & 0.39 & 0.59 & 0.42 & 0.28 & 0.37 & 0.33 & 0.40 & 0.75 & 6.50
\enddata
\end{deluxetable*}

In our simulations, we treat the three TrES telescopes as one single
instrument. This is clearly an oversimplification.  While these
telescopes are quite similar, there are differences between the data
quality and quantity obtained from each site.  For example, Sleuth
must contend with a higher sky background due to its proximity to
urban areas, whereas the fraction of clear nights for PSST is
typically lower than for the other two sites.  However, accounting for
these differences is difficult, and furthermore their effect on our
predicted detections is small compared to the other uncertainties in
our model. We therefore ignore these effects.  In dealing with the
window probability of the TrES network, we were greatly aided by the
plots posted on the Sleuth website which showed the window probability
for each field, as well as the total network observing time each field
had received.

We simulated thirteen different TrES fields using this
methodology. The network observation times and the location of each
field are shown in Table 2. The data for this table was collected off
from the Sleuth observing website. The expected detections for each of
the fields (and the total number of detections over all simulated
fields) are shown in Table 3. The effects of our Galactic structure
model can clearly be seen in distribution of expected detections. Both
the UMa0 and Crb0 fields are substantially above the plane of the
Galaxy ($b \approx 48^\circ$), and thus show a much lower detection
rate than those fields directed towards denser regions of the sky.

Figure 2 shows the distributions of detections over various parameters
in the Lyr1 field. For Figure 2, we show two distributions, one
assumes white noise in the photometry, the other assumes 3 mmag of red
noise. For the white noise estimates, we used a signal-to-noise limit
of S/N $\geq 30$ and a magnitude limit of $m_{R}\leq13$. Lyr1 is the
field in which TrES-2 was discovered, and is adjacent to Lyr0, which
contains TrES-1. We therefore have plotted the location of these two
stars on the distributions shown in Figure 2. Both stars reside close
to the peaks of our parent star mass and radius distributions. The
third and fourth planets discovered by TrES, TrES-3 and TrES-4, are
not plotted in Figure 2, as they reside in the Hercules fields, which
are at a higher Galactic latitude than Lyra. Also of note is that with
our assumptions half of the expected detections the F-dwarf
range. Among the false positives encountered by the TrES team, the
most frequent are eclipsing F/M-dwarf binaries. Figure 2 offers a
partial explanation for this, showing that roughly half of TrES'
planet-like signals come from F-dwarfs.

\begin{deluxetable}{|c|ccc|}
\tablecaption {\sc Predicted TrES Detections for $m_R\leq13$ and $R_p=1.1R_{Jup}$ }
\tablewidth{0pt}
\tabletypesize{\small}
\startdata
\hline
\hline
            &  VHJ &   HJ & Both\\
\hline
S/N$\geq20$ & 6.06 & 5.72 & 11.78\\
S/N$\geq25$ & 4.68 & 4.05 & 8.73\\
S/N$\geq30$ & 3.59 & 2.91 & 6.50\\
S/N$\geq35$ & 2.77 & 2.13 & 4.90\\
S/N$\geq40$ & 2.16 & 1.59 & 3.75
\enddata
\end{deluxetable}

\begin{deluxetable}{|c|ccc|}
\tablecaption {\sc Predicted TrES Detections for S/N$\geq30$ and $R_p=1.1R_{Jup}$  }
\tablewidth{0pt}
\tabletypesize{\small}
\startdata
\hline
\hline
              &  VHJ &   HJ & Both\\
\hline
$m_R\leq12.0$ & 2.13 & 1.95 & 4.08\\
$m_R\leq12.5$ & 2.95 & 2.58 & 5.53\\
$m_R\leq13.0$ & 3.59 & 2.91 & 6.50\\
$m_R\leq13.5$ & 3.86 & 3.01 & 6.87\\
$m_R\leq14.0$ & 3.94 & 3.03 & 6.97\\
$m_R\leq14.5$ & 3.96 & 3.05 & 7.01
\enddata
\end{deluxetable}

\begin{deluxetable}{|c|ccc|}
\tablecaption {\sc Predicted TrES Detections for S/N$\geq30$ and $m_R\leq13$ }
\tablewidth{0pt}
\tabletypesize{\small}
\startdata
\hline
\hline
                 &  VHJ &   HJ & Both\\
\hline
$R_p=0.9R_{Jup}$ & 1.66 & 1.17 & 2.83\\
$R_p=1.0R_{Jup}$ & 2.52 & 1.90 & 4.42\\
$R_p=1.1R_{Jup}$ & 3.59 & 2.91 & 6.50\\
$R_p=1.2R_{Jup}$ & 4.53 & 3.88 & 8.41\\
$R_p=1.3R_{Jup}$ & 5.52 & 5.04 & 10.56
\enddata
\end{deluxetable}

Our predicted number of detections for the selected TrES fields of
$N_{det} = 6.50$ covers only half of the fields observed by TrES, so
we may roughly extrapolate this result to $13.00$ detections over all
of the fields. This is roughly three times the four planets that TrES
has actually detected, though we hesitate to draw strong conclusions
from this comparison. While we have the luxury in our predictions of
imposing draconian signal-to-noise ratio and magnitude cuts in the
data, in real life the TrES detections are identified in data using
more flexible selection effects. A better comparison between our
predictions and the actual results of TrES (or any survey) would
therefore require a better understanding of the selection effects in
the observed data set.

\begin{deluxetable}{c|c}
\tablecaption {\sc XO Survey Parameters}
\tablewidth{0pt}
\tabletypesize{\small}
\startdata
\hline
\hline
Band & 400-700 nm\\
$m_V$ Limit & $12$\\ 
Observation Time & $\approx 400$ hrs\\
Exposure Time & 54 seconds\\
CCD Read Time & n/a\\
Telescope Diameter & 0.11 m (two)\\
Throughput & 0.16\\
PSF (FWHM) & 45''\\
$\theta_{pix}$ & 25.4''\\
CCD Full-Well Depth & $1.5\times10^5\ e^-$\\
\hline
\hline
Field Area & $7.2\times63\ \mathrm{deg}^2$\\
Galactic Longitude & Varies\\
Galactic Latitude & Varies
\enddata
\end{deluxetable}

To see how different selection criteria change our predictions for
TrES, Tables 4-6 show the effects of changing the S/N limit, the
magnitude limit, and the radius of the targeted planets on the
detection numbers.

A changing magnitude limit is particularly interesting, since all four
of the TrES detections have been around stars with $m_R\leq12$. This
might indicate that the effective magnitude limit of the TrES survey
is brighter than what is described in the published literature. If
this is the case, then using a limit of $m_R\leq12$ lowers the number
of expected detections in our simulated fields to $N_{det} = 4.08$,
and total detections to $8.16$.

\subsection{Drift-scanning - XO}

\begin{figure}[b]
\epsscale{1.2} 
\plotone{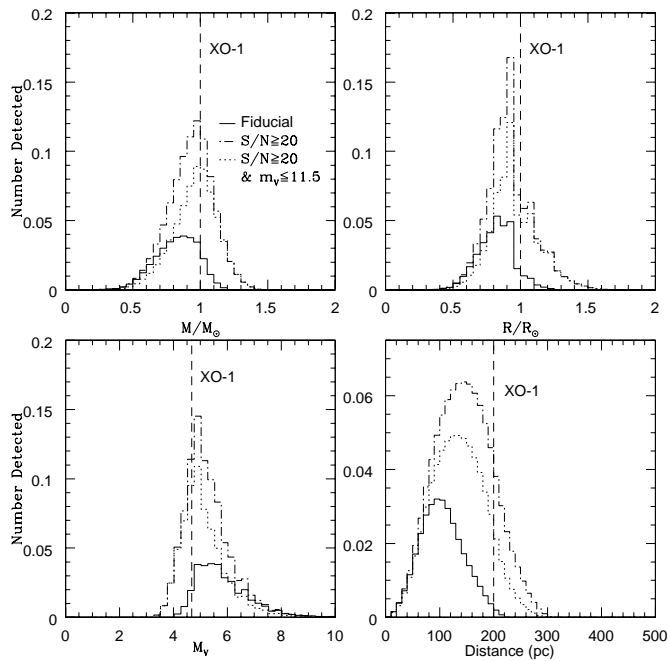}
\vskip -0.0in 
\figcaption[]{The distribution of predicted XO detections in the field at 16 hrs plotted against various parameters. Three cases are shown: our fiducial S/N$\geq30$, $m_V\leq12$ case, S/N$\geq20$, $m_V\leq12$, and S/N$\geq20$, $m_V\leq11.5$. Also shown is the actual location of the star XO-1, which is located within this field.}
\end{figure}

The XO survey is a wide-angle ground-based photometric survey that uses drift-scanning for its observations. The XO telescope is a pair of 11 cm cameras in place at the Haleakala Observatory in Hawaii. It monitors thousands of stars with $m_V\leq12$ in six fields spaced evenly around the sky in right ascension\footnote{00 hr, 04 hr, 08 hr, 12 hr, 16 hr, and 20 hr.}. The cameras scan within these fields, sweeping in declination from $0^\circ$ to $+63^\circ$ every ten minutes. Typically, this provides over 3000 observations per star in each of the XO fields. More on the specifics of the XO system can be found in \cite{mccullough2005}. The relevant parameters for our predictions are listed in Table 7.

To determine the throughput of XO and the background sky magnitude in the XO band (400-700 nm), we fit to the RMS diagram shown in Figure 8 of \cite{mccullough2005}. With the addition of their stated calibration and scintillation error, we were able to exactly reproduce the relations shown in that figure using a background sky magnitude of 19.4 mag $\mathrm{arcsec}^{-1}$ and a system throughput of 0.16.

\begin{deluxetable}{cccccc|c}
\tablecaption {\sc Predicted XO Detections for $m_V\leq12$, S/N$\geq30$, and $R_p=1.1R_{Jup}$ }
\tablewidth{0pt}
\tabletypesize{\small}
\startdata
\hline
\hline
00 hrs & 04 hrs & 08 hrs & 12 hrs & 16 hrs & 20 hrs & Total\\
\hline
  0.34 &   0.35 &   0.34 &   0.30 &   0.32 &   0.37 &  2.02
\enddata
\end{deluxetable}

Additionally, we checked to ensure that the values we used for the observation time and the number of nights observed generated an analytic window probability similar to that shown in Figure 7 of \cite{mccullough2005}. Our ability to do this was greatly aided by the XO team's description of how many nights they lost to weather and mechanical trouble.

We simulated each of the six XO fields for our predictions. The results for each field using the base parameters of S/N$\geq30$, $m_V\leq12$ and $R_p=1.1R_{Jup}$ are shown in Table 8. The first planet discovered by the XO survey, XO-1b, \citep{mccullough2006} is located in in the field at 16 hrs, and so we have also plotted the distribution of detections within this field over a variety of parameters in Figure 3. XO-2b \citep{burke2007} and XO-3b \citep{johnskrull2007} are in noticeably different fields (08 hrs and 04 hrs, respectively), so were not included on the plot.

\begin{deluxetable}{|c|ccc|}
\tablecaption {\sc Predicted XO Detections for $m_V\leq12$ and $R_p=1.1R_{Jup}$ }
\tablewidth{0pt}
\tabletypesize{\small}
\startdata
\hline
\hline
            &  VHJ &   HJ & Both\\
\hline
S/N$\geq20$ & 3.45 & 2.41 & 5.86\\
S/N$\geq25$ & 2.20 & 1.27 & 3.47\\
S/N$\geq30$ & 1.37 & 0.65 & 2.02\\
S/N$\geq35$ & 0.83 & 0.33 & 1.16\\
S/N$\geq40$ & 0.50 & 0.17 & 0.67
\enddata
\end{deluxetable}

\begin{deluxetable}{|c|ccc|}
\tablecaption {\sc Predicted XO Detections for S/N$\geq30$ and $R_p=1.1R_{Jup}$  }
\tablewidth{0pt}
\tabletypesize{\small}
\startdata
\hline
\hline
              &  VHJ &   HJ & Both\\
\hline
$m_V\leq11.5$ & 1.09 & 0.55 & 1.64\\
$m_V\leq12.0$ & 1.37 & 0.65 & 2.02\\
$m_V\leq12.5$ & 1.50 & 0.69 & 2.19\\
$m_V\leq13.0$ & 1.54 & 0.70 & 2.24\\
$m_V\leq13.5$ & 1.55 & 0.71 & 2.46
\enddata
\end{deluxetable}

\begin{deluxetable}{|c|ccc|}
\tablecaption {\sc Predicted XO Detections for S/N$\geq30$ and $m_V\leq12$ }
\tablewidth{0pt}
\tabletypesize{\small}
\startdata
\hline
\hline
                 &  VHJ &   HJ & Both\\
\hline
$R_p=0.9R_{Jup}$ & 0.29 & 0.09 & 0.38\\
$R_p=1.0R_{Jup}$ & 0.69 & 0.25 & 0.94\\
$R_p=1.1R_{Jup}$ & 1.37 & 0.65 & 2.02\\
$R_p=1.2R_{Jup}$ & 2.07 & 1.18 & 3.25\\
$R_p=1.3R_{Jup}$ & 2.93 & 1.92 & 4.85
\enddata
\end{deluxetable}

Tables 9-11 show the effects of changing the S/N limit, the magnitude limit, and the radii of the targeted planets on XO's predicted detection numbers. For our base case of $m_V\leq12$ and S/N$\geq30$, we expect that XO will detect 2.02 planets. If we consider the three actual detections from XO, it seems that, similar to TrES, the actual S/N and magnitude limit of the XO survey may differ from what is in the literature. For example, all three XO parent stars are at $m_V\leq11.5$ (XO-1, the dimmest, is at $m_V=11.3$). To see how this affected our calculated distributions, we also simulated the field at 16 hrs assuming S/N$\geq20$, $m_V\leq12$, and S/N$\geq20$, $m_V\leq11.5$. The results are included on Figure 3. In these cases, the total number of detections increases to 5.86 and 4.21, respectively.

\subsection{Space-based - Kepler}

To make predictions for a space-based mission, we chose NASA's Kepler mission, which is primarily designed to look for Earth-sized extrasolar planets. It consists of a single spacecraft that is currently scheduled to be launched into an Earth-trailing heliocentric orbit in early 2009. Onboard, Kepler will have 42 CCDs that will measure the light from 100,000 main-sequence stars within the telescope's field of view. Ideally, Kepler will identify planets orbiting within the ``habitable zone'' of their parent stars which would be capable of possessing liquid water on their surfaces.

\begin{deluxetable}{c|c}
\tablecaption {\sc Kepler Mission Survey Parameters}
\tablewidth{0pt}
\tabletypesize{\small}
\startdata
\hline
\hline
Band & 400-850 nm\\
$A_{Kepler}$ & 0.86\\
$m_V$ Limit & $\approx 14$\\ 
Observation Time & 4 years\\
Exposure Time & 3 seconds\\
CCD Read Time & Negligible\\
Telescope Diameter & 0.95 m\\
Throughput & .3\\
PSF (FWHM) & 10''\\
$\theta_{pix}$ & 3.98''\\
CCD Full-Well Depth & $1.2\times10^6\ e^-$\\
\hline
\hline
Field of View & $106\ \mathrm{deg}^2$\\
Galactic Longitude (new) & $76.32^\circ$\\
Galactic Longitude (old) & $70^\circ$\\
Galactic Latitude (new) & $13.5^\circ$\\
Galactic Latitude (old) & $5^\circ$
\enddata
\end{deluxetable}

Beyond looking for extrasolar Earths, the immense amount of photometry that Kepler is expected to produce over its four year operational lifetime will also allow for the detection of larger jovian worlds transiting other stars. Previous estimates of the number of VHJs and HJs expected to be found by Kepler have predicted the discovery of about 180 HJs (no VHJs had been detected at the time, and so they were not treated) \citep{jenkins2003}. Clearly, given that there are only 46 currently known transiting planets, this would be a significant increase in the number of known planets

To simulate Kepler with our methodology, we used the values in Table 12 to simulate Kepler observations out to a distance of $8\ \mathrm{kpc}$, using a signal-to-noise ratio of SN$\geq7$. We gathered the information in Table 12 from both the Kepler mission website\footnote{http://kepler.nasa.gov/} as well as published descriptions of the satellite \citep{borucki2004}. Three items are of note. First, our value for the photon detection efficiency of Kepler is based upon information from the Kepler website that a 12th magnitude G2 dwarf will show $7.8\times10^8\ \gamma\  \mathrm{hr}^{-1}$. Secondly, the value we use for the extinction ratio in the Kepler bandpass, $A_{Kepler}/A_V = 0.861$, is an average that we calculated by integrating the extinction curve over the Kepler bandpass (400-850nm) \citep{fitzpatrick1999}. Finally, the Kepler website and much of the literature states that Kepler will use a dim magnitude cutoff of $m_V\leq14$. In reality, there is no ``hard'' magnitude cutoff at which stars will be excluded from observations. Indeed, Kepler is expected to observe promising stars that are much dimmer than $m_V=14$ (D. W. Latham, private communication). Therefore, for our more explicit calculations of the expected number of HJ and VHJ detections for Kepler, we adjusted the magnitude limit such that the number of stars in the field agreed with the expected number of Kepler targets (100,000 in the case of the new field). For other, more general, simulations, we used several different magnitude limits to determine what effects these had on Kepler's expected detections.

While the location of the Kepler field is currently centered on Galactic coordinates $(l=76.32^\circ, b=13.5^\circ)$, up until 2004 it was instead situated closer to the Galactic plane at $(l=70^\circ, b=5^\circ)$. The switch in the position of the field to higher Galactic latitudes was done in order to avoid sorting through the larger number of giant stars that are found in fields directed along the plane of the Galaxy. Much of the previously published literature examining the Kepler mission have used the coordinates of the old field. For the purpose of comparison, we therefore simulated both the old and new Kepler fields.    

\subsubsection{General Characterisation}

\begin{figure}
\vskip -0.15in
\epsscale{1.2} 
\plotone{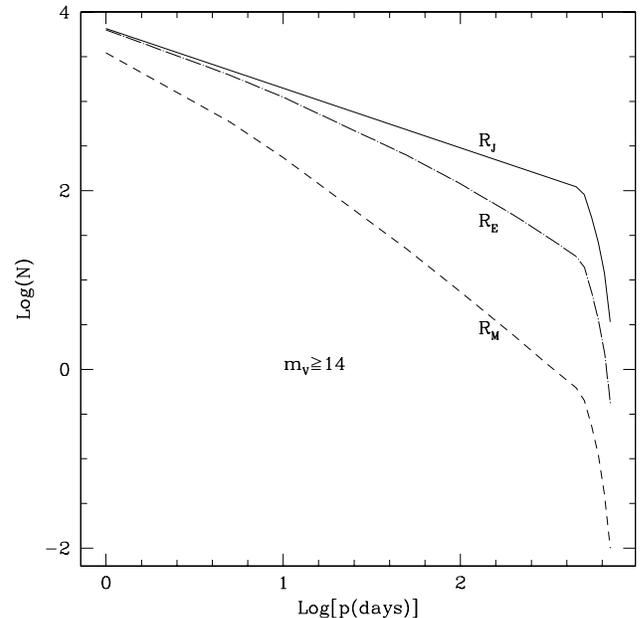}
\vskip -0.15in 
\figcaption[]{$\log(N)$ vs. $\log(p)$ for Kepler detections down to $m_V=14$ assuming that every star possess a planet, for Jupiter-, Earth-, and Mars-sized worlds. Note that both the Jupiter- and Earth-sized detection numbers are similar for $p\leq10$ days; this implies that in this radius and period regime the number of Kepler detections is limited by the magnitude cutoff, and not the S/N (as is the case for the Mars-size planets). The strong downturn around $\log(p)=2.7$ is a result of Kepler's window probability (i.e. the probability of seeing two or more transits).}
\end{figure}

Looking in the new Kepler field with a magnitude limit of $m_V\leq14$, we simulated the number of detections as a function of orbital period for Jupiter-, Earth-, and Mars-sized worlds, assuming that every star possessed a planet with the given period and radius (Figure 4), and that red noise was negligible. Interestingly, the difference between the number Jupiter- and Earth-sized planets detected is minimal for $p \leq 10$ days, which indicates that Kepler is able to detect all the planets within this radius range, assuming that they transit. Therefore, for $m_V \leq 14$ and $p \leq 10$ days, the S/N of transiting Jupiter- and Earth-sized planets is large enough that Kepler's S/N limit is irrelevant, and the $m_V\leq14$ magnitude cutoff is the limiting factor in the number of detections. For smaller planets the size of Mars, the S/N of their transits are in the neighborhood of the Kepler S/N limit for the brightest stars and shorter periods, and so Mars-sized worlds are detected at a much lower rate.

\begin{figure}
\vskip -0.0in
\epsscale{1.2} 
\plotone{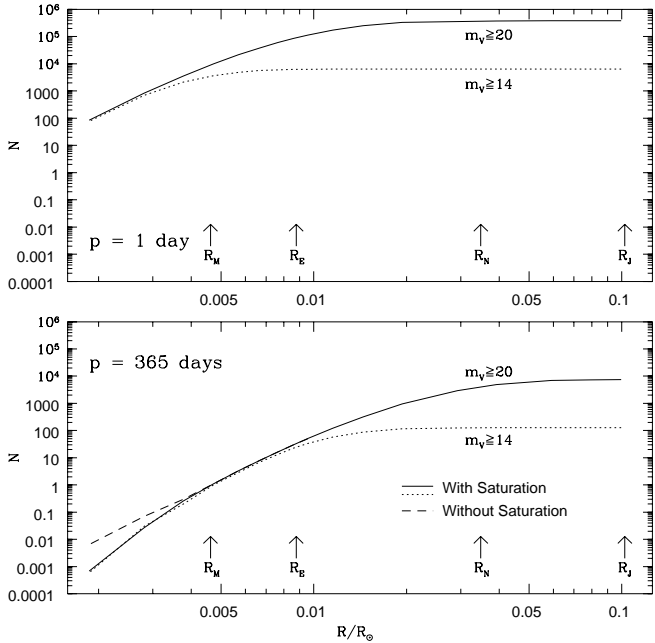}
\vskip -0.1in 
\figcaption[]{Kepler detections as a function of the planetary radius for magnitude limits of $m_V\geq14$ and $m_V\geq20$ assuming that all stars have planets. Also shown are markers for the radii of Jupiter, Neptune, Earth, and Mars. One can see that as the period increases, the planetary radius at which Kepler becomes magnitude-limited (the flat portions of the graphs) increases, while the overall number of detections falls.}
\end{figure}

To more accurately characterize the magnitude- and S/N-limited detection regimes, we plotted the number of detections as a function of planetary radius for orbital periods of 1 day and 365 days (Figure 5). In addition in a magnitude limit of $m_V=14$, we also plot the same relations for a limit of $m_V=20$. One can clearly see the flattening of these curves as the radius of the target planets increase, indicating the increasing effect of the magnitude limit on the detection numbers. As is expected in the S/N limited regime, detections go as $R_p^6$ \citep{pepper2003, gaudi2005, pepper2005}, though at the lower end saturation of the Kepler CCDs serves to cut out some of these detections. 

\subsubsection{Habitable Planets}

One of the major goals of the Kepler mission is the detection of planets in a star's habitable zone that would be capable of supporting life. The habitable zone is defined as the distance from the parent star at which liquid water water could exist on a planet's surface. We estimate the detection rate of habitable zone planets by assuming that all stars have a planet with a semi-major axis equal to
\begin{equation}\label{eq:415}
\frac{a}{{\rm AU}} = \sqrt{\frac{L_{bol,*}}{L_{bol,\odot}}}.
\end{equation}

This assumption is simple and likely optimistic.  In reality, the habitable zone has a finite width, and habitable planets will have a range of radii.  Thus, a better estimate would be to adopt a distribution of planet masses and radii, and integrate over the range of radii and periods considered habitable.  Furthermore, the assumption of one habitable planet per star is probably optimistic, and so the actual number of detected habitable Earths will be smaller than we estimate,  depending on the real frequency of Earth-like planets.  For the sake of simplicity, however, we will adopt the assumptions above.

With this assumption, we simulated the detection of habitable Earths in both the old and new Kepler fields (Figure 6). Surprisingly, despite there being nearly twice as many main sequence stars in the old field, the number of detected habitable Earths remained nearly the same in the two fields. This occurred because at the magnitude relevant for Kepler, the increased star count in the old field was mainly due to the larger number of A and F main-sequence stars found closer to the Galactic plane; habitable planets are doubly hard to detect around these stars due to the lessened depth of transit, and the longer periods required for a planet to be in the habitable zone. The exact distribution of stars by spectral type for the old and new fields is given in Table 13, and is discussed further below. For the 106,000 stars in the new Kepler field, we find that Kepler will be able to detect 79 habitable Earth radius planets, assuming that every star posses one. 

\begin{figure}
\vskip -0.125in
\epsscale{1.23} 
\plotone{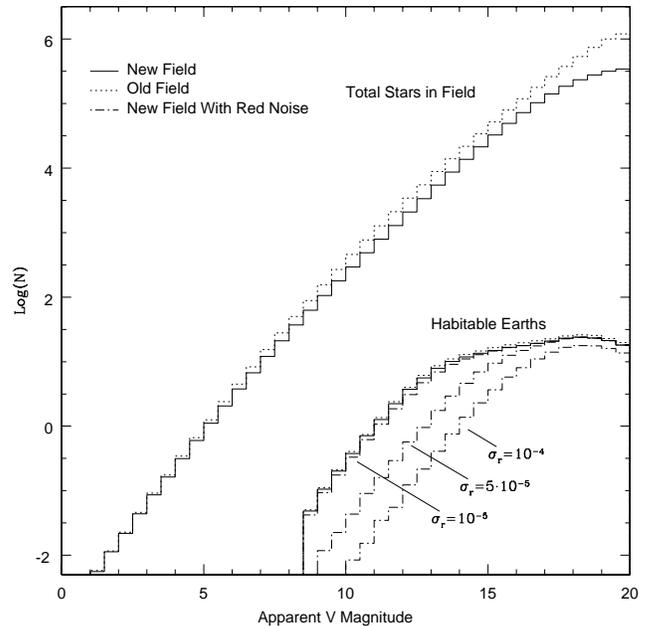}
\vskip -0.15in 
\figcaption[]{The number of habitable Earths detected by Kepler in the old and new fields binned by the apparent V-magnitude of the parent star, as well as the results assuming varying amounts of red noise on the new field. Also shown are the star counts for both fields binned in a similar manner.}
\end{figure}

In general, Figure 6 provides an excellent example of the benefit of repositioning the Kepler field away from the Galactic plane. Namely, the number of large dwarfs and giant stars (which we have not treated) found in the Galactic plane can be avoided, reducing the false-positive rate, while the number of habitable Earths detected remains nearly the same \citep{gould2003b}.

It is also possible that life may arise on non-Earth-sized worlds. A world the size of Mars might develop quite differently than it has in our own Solar System if it were closer to its parent star, and Jovian worlds in the habitable zone could possess large moons capable of supporting life. Figure 7 shows the number of detections of Jupiter-, Earth-, and Mars-sized habitable worlds, assuming that every star has a planet, against a variety of parameters in the new Kepler field. For completeness, the magnitude limit was set to $m_V\geq 20$. One can see in the mass and radius plots an explicit demonstration of why changing the number of bright, massive main-sequence stars in the field does not effect the number of habitable planet detections. As was mentioned earlier all detections drop to zero beyond $1.4\ M_\odot$. In the plot showing the number of detections binned against the absolute V-magnitude of the parent star we have additionally shown the effect of changing the apparent V-magnitude limit on the distribution of Earth-sized planets. 

\subsubsection{Hot and Very Hot Jupiters}

Using the HJ and VHJ frequencies provided by \cite{gould2006b}, we were able to make explicit predictions about the number of these worlds that Kepler should find. As mentioned previously, \cite{jenkins2003} (hereafter JD03) have made their own estimates of Kepler's HJ detection ability (as noted previously, no VHJs were known at the time, and so were not considered), thus providing a useful point of comparison. As before, we used a limit of S/N$\geq7$, though for Jupiter-sized planets Kepler is magnitude, and not signal-to-noise, limited.

\begin{figure}[b]
\vskip -0.0in
\epsscale{1.2} 
\plotone{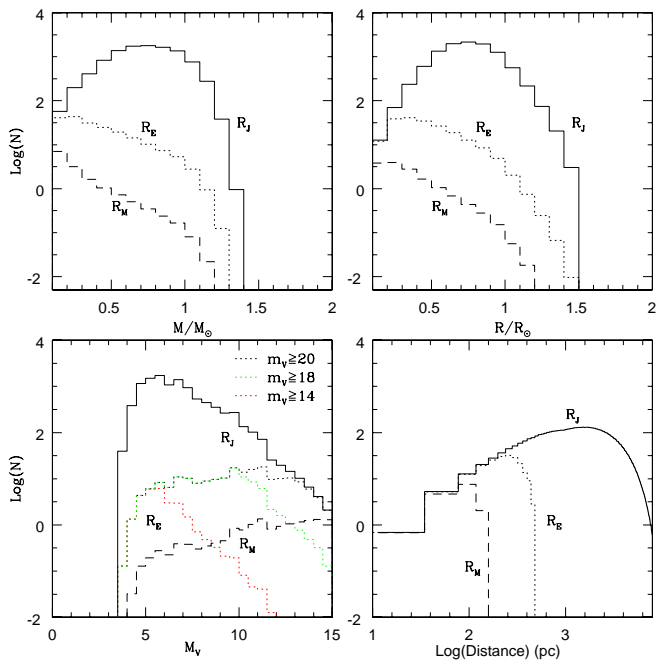}
\vskip -0.0in 
\figcaption[]{The number of Kepler detections of Jupiter-, Earth-, and Mars-sized worlds binned against a variety of parameters. These plots assume that every star possesses a planet, and use an apparent V-magnitude limit of $m_V\geq 20$. In the plot showing the distribution by absolute V-magnitude of the parent star, we have also plotted distributions showing the effect of changing the apparent magnitude limit on the distribution of detected Earth-sized worlds.}
\end{figure}

\begin{deluxetable*}{ccccccc|c}
\tabletypesize{\small}
\tablecaption{\sc Number of Main-Sequence Stars in Kepler's Field of View}
\tablewidth{0pt} 
\startdata
\hline
\hline
 & B & A & F & G & K & M & All\\  
\hline
      Jenkins \& Doyle: & 11014&  36708&  81085&  61347&  15573&   831&  206558\\
Calculated (old field): &  4278&  31370&  76630&  62891&  18243&  1264&  194678\\
Calculated (new field): &   338&   5066&  41094&  45568&  13351&   942&  106342
\enddata
\end{deluxetable*}

\begin{deluxetable*}{ccccccc|c}
\tabletypesize{\small}
\tablecaption{\sc Expected Transiting HJs and VHJs in Kepler's Field of View}
\tablewidth{0pt} 
\startdata
\hline
\hline
 & B & A & F & G & K & M & All (HJs)\\  
\hline
      Jenkins \& Doyle: &   10&    32&    71&    54&   14&     1&   181 (181)\\
Calculated (old field): &  4.2&  27.3&  55.5&    33&    8&   0.4&   127 (72.8)\\
Calculated (new field): &  0.3&   4.4&  28.6&  23.5&  5.8&  0.33&  62.9 (36)
\enddata
\end{deluxetable*}

At the time of the JD03 estimate the Kepler field was centered on its old coordinates. For the purposes of comparison, we therefore examined the old Kepler field in addition to the new field. 

Tables 13 and 14 show the distribution of our predictions by spectral type. To facilitate comparison with JD03, we have included not only the final prediction numbers for both old and new Kepler fields, but also the distribution of the calculated total number of stars in each field. As stated above, in both fields we set the magnitude limit of the simulation such that the total number of stars in the fields agreed with the number of expected Kepler targets. In the case of the new field, this is 100,000 stars, while for the old field we matched the number of stars predicted by JD03. This corresponded to a magnitude limit of $m_V=15.7$ in the new field and $m_V=15.9$ in the old field.

Looking at the detections, one can see in Table 14 that our final prediction of $N_{det}=127$ for the old Kepler field is lower than the JD03 prediction. This is primarily a result of the fact that the \cite{gould2006b} results that we use for $k(p)$ (see Section 2.2) posit a HJ and VHJ frequency approximately $1/3$ of the frequency used in JD03. An important difference, however, between our work and JD03 is our consideration of VHJs, which were not treated by JD03. This has the effect of ``doubling'' the number of planets we expect Kepler to detect (as demonstrated by the final values in parentheses in Table 14). We therefore arrive at detection rates for VHJs and HJs that are 2/3 that of JD03 for the old Kepler field, and 1/3 that of JD03 in the new Kepler field. For the old field, we predict 54.2 VHJs and 72.8 HJs. For the new field, we predict 26.9 VHJs, and 36.0 HJs. 

In the new Kepler field, we see that looking away from the plane of the Galaxy reduces the number of HJs and VHJs detected by half to $N_{det}=62.9$, which matches the overall drop in the star count in the new field. The distribution of detections changes substantially, shifting more towards F-, G-, and K-dwarfs. This is the caused by our model of the Galaxy's vertical structure, which uses reduced scale heights for earlier stars. Thus, most of the difference in star counts between the old and new fields is caused by fewer B-, A-, and F-dwarfs, leading to proportionally smaller detection numbers around these types of stars.  

While the prediction number of detections in the new Kepler field is lower than the corresponding JD03 prediction would be for the same star count, this is accounted for in the same way as in the old field. Indeed, we find that our predictions are consistent with those of JD03, except for the difference in our assumption about the frequency of HJs and VHJs. The eventual results from Kepler therefore offer an excellent opportunity to more robustly determine the actual prevalence of short period Jovian worlds.

\subsection{All-sky Surveys}

\begin{deluxetable*}{c|cccc}
\tablecaption {\sc All-Sky Survey Parameters}
\tablewidth{0pt}
\tabletypesize{\small}
\startdata
\hline
\hline
Survey                 & LSST     & SDSS-II   & Pan-STARRS & Pan-STARRS Wide\\
\hline
$m_V$ Limit (Sun-like) & 18.51    & 15.56     & 14.99      & 12.49\\ 
$m_V$ Limit (M-dwarfs) & 23.13    & 20.18     & 19.61      & 17.11\\ 
Observation Time       & 10 years & 37.5 days & 5 months   & 5 months\\ 
Telescope Diameter     & 6.5 m    & 2.5 m     & 1.8 m      & 1.8 m\\ 
Throughput             & 0.5      & 0.5       & 0.5        & 0.5\\ 
\hline
\hline
Field of View & $9.6\ \mathrm{deg}^2$    & $6.25\ \mathrm{deg}^2$ & $7\ \mathrm{deg}^2$ & $7\ \mathrm{deg}^2$\\
Area Surveyed & $20,000\ \mathrm{deg}^2$ & $300\ \mathrm{deg}^2$  & $1200\ \mathrm{deg}^2$ & $12,000\ \mathrm{deg}^2$
\enddata
\end{deluxetable*}
In addition to wide-field photometric surveys, several current and planned projects aim to repeatedly gather photometry over gigantic swathes of the sky. While not all of them have extrasolar planet detection as one of their goals, the large areas covered by these surveys, together with their long durations (from 3 to 10 years), offer the possibility of detecting hundreds of transiting VHJs and HJs.

We examined three all-sky surveys, LSST, SDSS-II, and Pan-STARRS. The Large Synoptic Survey Telescope (LSST) will be a large 8.4 m telescope, with an effective aperture of 6.4 m, situated in northern Chile. It is scheduled to finish construction sometime in 2014. Using a 9.6 deg$^2$ field of view, the LSST team intends to image 20,000 deg$^2$ of the sky repeatedly over the course of 10 years.\footnote{http://www.lsst.org} 

SDSS-II is the second phase of the Sloan Digital Sky Survey (SDSS) which began in June, 2005 and is scheduled to last for three years until June of 2008. It is composed of three distinct surveys, the Sloan Legacy Survey, SEGUE, and the Sloan Supernova Survey \citep{sako2005}. The Supernova Survey is the only one of these three that will collect photometry capable of finding transiting planets; the other two survey modes are focused on non-variable objects within and outside the Galaxy. To execute the Supernova Survey, SDSS uses a 2.5 m telescope on Apache Point in New Mexico with a 6.25 deg$^2$ field of view. Over the three year run of SDSS-II, the team expects the Supernova Survey to image about 300 deg$^2$ of the sky near the Galactic poles every two nights for the three months a year the poles are visible. Given the observing cadence the survey expects to use, this translates into a total observing time of 37.5 days.

\begin{deluxetable}{|c|ccc|}
\tablecaption {\sc Transiting VHJ and HJ Densities Around Sun-like (G2) Stars (deg$^{-2}$) }
\tablewidth{0pt}
\tabletypesize{\small}
\startdata
\hline
\hline
Mag. Limit               & Gal. Plane & Gal. Poles & All-Sky Average\\
\hline
$m_V \leq 20$            & 4.052      & 0.027      & 0.800\\
LSST (18.51)             & 1.609      & 0.027      & 0.387\\
$m_V \leq 18$            & 1.125      & 0.026      & 0.293\\
$m_V \leq 16$            & 0.219      & 0.025      & 0.087\\
SDSS-II (15.56)          & 0.146      & 0.020      & 0.063\\
Pan-STARRS (14.99)       & 0.083      & 0.016      & 0.041\\
$m_V \leq 14$            & 0.029      & 0.009      & 0.017\\
Pan-STARRS Wide (12.49)  & 0.005      & 0.003      & 0.004\\
$m_V \leq 12$            & 0.003      & 0.002      & 0.002
\enddata
\end{deluxetable}

\begin{deluxetable}{|c|ccc|}
\tablecaption {\sc Transiting VHJ and HJ Densities Around M-dwarfs (M5) (deg$^{-2}$) }
\tablewidth{0pt}
\tabletypesize{\small}
\startdata
\hline
\hline
Mag. Limit               & Gal. Plane & Gal. Poles & All-Sky Average\\
\hline
LSST (23.13)             & 2.7558     & 0.0888     & 0.7764\\
SDSS-II (20.18)          & 0.2473     & 0.0398     & 0.1136\\
$m_V \leq 20$            & 0.2081     & 0.0368     & 0.0989\\
Pan-STARRS (19.61)       & 0.1422     & 0.0304     & 0.0725\\
$m_V \leq 18$            & 0.0257     & 0.0105     & 0.0169\\
Pan-STARRS Wide (17.11)  & 0.0092     & 0.0048     & 0.0068\\
$m_V \leq 16$            & 0.0047     & 0.0015     & 0.0028\\
$m_V \leq 14$            & 0.00017    & 0.00015    & 0.00016\\
$m_V \leq 12$            & 0.00001    & 0.00001    & 0.00001
\enddata
\end{deluxetable}

Finally, Pan-STARRS - the Panoramic Survey Telescope and Rapid Response System - will be a group of four 1.8 m telescopes atop the summit of Haleakala, Hawaii \citep{kaiser2004}. The first telescope, PS-1, is nearing completion. When fully operational, Pan-STARRS will embark on a 3 year-long ``Medium Deep'' survey that will observe 1200 square degrees of the sky near the Galactic poles for a total of five months.\footnote{The remaining balance of the 3 years will be taken up by the other Pan-STARRS science modes; see \\ http://pan-starrs.ifa.hawaii.edu/project/reviews/PreCoDR/...\\ documents/PSCoDD\_1\_4\_design\_reference\_mission.pdf} We also looked at a notional ``wide'' Pan-STARRS survey that looked at 10x the amount of sky as the Medium Deep survey, but used the same amount of observing time.

To characterise each of these surveys and find how many transiting VHJs and HJs they should discover, we first calculated the V-magnitude limits at which each of the surveys will be able to detect transiting Jovian worlds. The complete derivation of this limit is given in Appendix E. Table 15 shows the results of these calculations, as well as the other relevant parameters for each of the surveys. We calculated magnitude limits for detecting giant planets around both Sun-like ($0.8 \leq M_*/M_\oplus \leq 1.2$) stars and M-dwarfs ($0.1 \leq M_*/M_\oplus \leq 0.5$) for S/N$\geq20$.

\begin{figure}[b]
\vskip -0.1in
\epsscale{1.2} 
\plotone{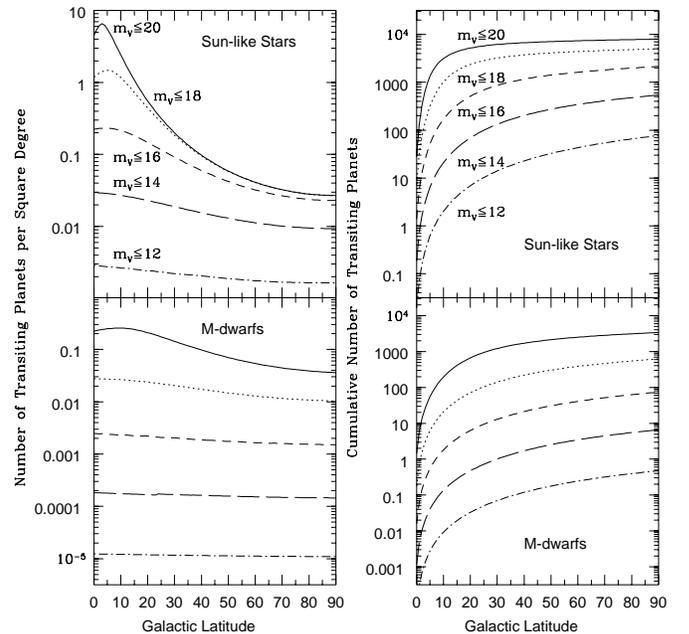}
\vskip -0.1in 
\figcaption[]{The number of transiting planets per square degree around Sun-like stars and M-dwarfs for several V-magnitude limits. The left plots are differential, showing the value at a particular Galactic latitude. The right plots are cumulative, showing the total planets from zero degrees up to a particular galactic latitude.}
\end{figure}

To translate these magnitude limits into the number of transiting planets each of the surveys should detect, we used our methodology to calculate the density of VHJs and HJs for various limiting magnitudes, in numbers per square degree, along the Galactic plane, at the Galactic poles, and averaged around the sky. Table 16 shows the resultant values for a series of fiducial magnitude limits, as well as the planet densities at the magnitude limits of our modeled surveys for Sun-like stars; Table 17 shows transiting planet densities around M-dwarfs. Additionally, in Figure 8 we show the transiting planet density around Sun-like stars and M-dwarfs as a function of Galactic latitude. It is interesting to note the dip near the Galactic plane, which occurs as a result of extinction from interstellar dust. 

Knowing the angular size of each survey, we use these densities to predict the number of planets that will be seen. For SDSS-II and Pan-STARRS, which are both pointed towards the Galactic poles, we used the polar densities, while for LSST and the wide Pan-STARRS, with their much larger and wide-ranging survey area, we used the all-sky average density. The final predictions for both Sun-like stars and M-dwarfs are shown in Table 18.

We note that these predictions are based on a simpler model than what we have used in other portions of this paper. For instance, we do not include the effects of CCD saturation or sky-background. Nevertheless, our calculations should provide a good expectation for the range of the detection numbers that these all-sky surveys can expect. 

The large difference between the number of planets expected to be detected by LSST and the other surveys is caused by two effects. First, SDSS-II and Pan-STARRS cover much smaller areas of the sky to shallower depths than LSST, and second, LSST is not primarily directed towards the Galactic poles, greatly increasing the density of stars for which planet detection is possible. Pan-STARRS Wide, meanwhile, has a substantially brighter magnitude limit than any of the other surveys, which prevents it from detecting as many planets as LSST, despite the similarities in the size of their survey areas. Pan-STARRS Wide does detect more planets than the planned Narrow survey, primarily because the latter ``runs up'' against the scale height of the Galaxy.

\begin{deluxetable}{c|cc}
\tablecaption {\sc All-Sky Number of VHJs and HJs Detected }
\tablewidth{0pt}
\tabletypesize{\small}
\startdata
\hline
\hline
                & Sun-like (G2) & M-dwarfs (M5)\\
\hline
LSST            & 7740     & 15530\\
SDSS-II         & 6.0      & 11.9\\
Pan-STARRS      & 19.2     & 36.5\\
Pan-STARRS Wide & 48.0     & 81.6
\enddata
\end{deluxetable}

The real difficulty, however, in managing the 7,740 or 15,530
detections that we expect LSST to make will be properly identifying
and separating the false-positives from the actual planetary transit
events. Indeed, to date the wide-angle surveys have generated
approximately 15 false positive candidates for every actual
transiting planet they have seen (D. W. Latham, private
communication). If we extend this ratio to LSST, SDSS-II, Pan-STARRS
and Pan-STARRS Wide, we would expect them to create 116,000, 90, 290,
and 720 false positives, respectively around Sun-like stars, and
233,000, 180, 550, and 1,200 around M-dwarfs, respectively. Working
through this many transit candidates (especially in the case of LSST)
is an intensive, time-consuming task. We would therefore expect the
all-sky surveys, despite the high absolute number of predicted
detections, to discover transiting planets at about the same rate as
the operational wide-angle surveys, as the identification rate is
predominantly limited by the ability to make follow-up observations of
candidates to determine whether or not it is a false positive.

\subsection{Effects of Red Noise}

As pointed out by \cite{pont2006}, the general mathematical description we have used for the S/N of a transit, equation (\ref{eq:30}), assumes uncorrelated errors on the collected photometric data points (white noise). In practice this may not be the case, as phenomenon such as changes in the seeing, airmass, telescope tracking, etc. introduce systematic trends into the observed data that are correlated on the timescales of transits (red noise). This has the effect of substantially decreasing the effective S/N displayed by a transit, thus making detections more difficult. Specifically, \cite{pont2006} show that to account for red noise, equation (\ref{eq:30}) must be modified by adding a term to include a covariance coefficient $C_{ij}$ between the $i^{th}$ and $j^{th}$ measurements,
\begin{equation}\label{eq:600}
\chi^2 = \delta^2 \left( \frac{\sigma^2}{N_{tr}} + \frac{1}{N_{tr}^2} \displaystyle\sum_{i\neq j} C_{ij}\right)^{-1}.
\end{equation}       
This is equation (4) in \cite{pont2006}, using the variables we previously defined for our equation (\ref{eq:30}). The authors give a more manageable version on page 8 of their work, which relates the S/N with red noise to the S/N without red noise:
\begin{equation}\label{eq:610}
\chi_r^2 = \chi_w^2\ \left[ 1 + N_k \left(\frac{\sigma_r}{\sigma}\right)^2\right],
\end{equation}
where $\chi_w^2$ is the white noise $\chi^2$ statistic we calculate in equation (\ref{eq:30}), $N_k$ is the number of data points in the single $k^{th}$ transit (which on average is just $N_{tr}/$number of transits), and $\sigma_r$ is the red component of the noise in the data.

\begin{deluxetable*}{c|ccccccccccccc|c}
\tablecaption {\sc Predicted TrES Detections With 2 mmag Red Noise, S/N$\geq12$ }
\tablewidth{0pt}
\tabletypesize{\scriptsize}
\startdata
\hline
\hline
     & And0 & Cyg1 & Cas0 & Per1 & UMa0 & CrB0 & Lyr1 & And1 & And2 & Tau0 & UMa1 & Her1 & Lac0 & Total\\
\hline                                                                            
VHJs & 0.34 & 0.33 & 0.31 & 0.30 & 0.17 & 0.20 & 0.30 & 0.22 & 0.17 & 0.20 & 0.16 & 0.20 & 0.36 & 3.26\\
HJs  & 0.25 & 0.22 & 0.21 & 0.21 & 0.11 & 0.13 & 0.19 & 0.12 & 0.07 & 0.11 & 0.11 & 0.14 & 0.25 & 2.15\\
\hline                                                                            
Both & 0.59 & 0.55 & 0.52 & 0.51 & 0.28 & 0.33 & 0.49 & 0.34 & 0.24 & 0.31 & 0.27 & 0.34 & 0.61 & 5.41
\enddata
\end{deluxetable*}

\begin{table*}[t]
\centering
\begin{tabular}{c}
\end{tabular}
\end{table*}

To properly use this formulation, we must therefore have some idea of
what value $\sigma_r$ takes. \cite{pont2006, pont2007} infer values of 3 mmag of red noise for typical ground-based surveys, which can be reduced to approximately 2 mmag after applying de-trending algorithms to the data.  We therefore
took this tentative value of $\sigma_r = 0.002$ and used it to
re-simulate the TrES fields to see what effect red noise would have on
detection numbers. As noted in \cite{pont2007}, we set a signal-to-noise ratio limit of $S/N_r \geq 12$, which is the $S/N_r$ value of the detection of TrES-1b; TrES-2b has a higher value of $S/N_r = 14$. The results are shown in Table 19. Figure 2 shows
(in addition to TrES with only white noise) the distribution of
detections using 2 mmag of red noise and $S/N_r \geq 12$.
      
\begin{deluxetable}{c|c}
\tabletypesize{\scriptsize}
\tablecaption{\sc Kepler and Red Noise}
\tablewidth{0pt} 
\startdata
\hline
\hline
$\sigma_r$ & Hab. Earths\\
\hline
$10^{-4}$ & 14.37\\
$5\cdot 10^{-5}$ & 32.53\\
$10^{-5}$ & 72.88\\
0 & 78.89
\enddata
\end{deluxetable} 

Compared to a white S/N limit of $S/N_w \geq 30$, using $\sigma_r = 2$
mmag and $S/N_r \geq 12$ drops the total number of expected detections
by about 20\%. Looking at the split between HJs and VHJs,
the HJ population is disproportionately effected, which is a result of
the relatively fewer transits HJs show, lowering the overall S/N of
their transit signal. Similarly, Figure 2 shows that red noise also
cuts out the detections around higher mass stars, as these have a
lower S/N overall. 

Without de-trending the data, a red noise of 3 mmag would cause an even further reduction in detections, down 60\% as compared to white noise.

We also examined the possible effect that red
noise would have on the yield of habitable Earth-radius planets from
Kepler mission.  Given that Kepler is a
space-based mission designed to detect transit signals with fractional
depths at the level of
$\sim 10^{-4}$, we would expect that the amount of red noise due to
instrumental and environmental sources would be small, if it exists at all.  However,
this is generally difficult to predict before launch.  Furthermore,
there may exist significant astrophysical red noise due to intrinsic variability.  

We therefore estimated the expected Kepler yield for
a set of three values for $\sigma_r$, assuming that every star has an Earth-radius planet
with $a={\rm AU}(L/L_\odot)^{1/2}$ (Table 20).
For a relatively large amount of red noise at $\sigma_r =
10^{-4}$, we find the number of
habitable Earths-radius is drastically reduced from our baseline (uncorrelated noise) prediction,
from 78.89 down to 14.37. This is not surprising, since Earth passing
in front of a Sun-like star has a transit depth of only $8.5\cdot
10^{-5}$. For this amount of red noise, Kepler is
restricted to detecting Earth-like planets only around K- and
M-dwarfs. This shifts the expected distribution of host stars towards
dimmer magnitudes (Figure 6).
For smaller values of $\sigma_r = 5\cdot10^{-5}$ and $\sigma_r =
10^{-5}$, the number of habitable Earths detected is reduced to 32.53
and 72.88, respectively. Both of these estimates also assume that
every target star has an Earth-sized planet in the habitable zone, and
therefore must be viewed as upper limits. 

Thus we conclude that Kepler's yield of habitable zones of habitable
planets will be not be strongly affected by red noise, provided that
the red noise is $\la 10^{-5}$ on the timescales of the transits,
which is $\sim 13~{\rm hrs}(M/M_\odot)^{1.75}$ for solar-type stars.  If
the red noise is found to be larger than this, it may
compromise the primary Kepler science by significantly reducing the number
of expected habitable planet detections.  For example, if the red
noise is $\sigma_r = 10^{-4}$, and the frequency of habitable planets
with $r=R_\oplus$ is $\sim 20\%$, there is a $\sim 5\%$ chance that
Kepler will not detect any.  The effects of red noise on the habitable
planet yield can be mitigated by including a larger number of K and M
dwarfs in the target sample.  Of course, it will also be possible to
extrapolate the frequency of shorter-period or more massive planets, where the
effects of red noise are mitigated, into the habitable zone.

\section{Conclusion}

In this paper we have developed a method to more accurately predict the number of transiting planets detectable by wide-field ground- and space-based photometric surveys. We have taken into account such factors as the frequency of gas giants around main sequence stars, the probability of transits, stellar density changes from Galactic structure, and the effects of interstellar extinction. We adopt a S/N detection threshold criterion to approximate the detectability of planetary transits. 

To test our model, we compared our predictions to two operating photometric surveys, TrES and XO. For both, we calculated the number of statistically expected detections, as well as the population characteristics of the host stars. For TrES, we overestimate the number of detections (thirteen versus the four actual detections), while for XO our simulations yield approximately the actual number of detections (two predicted versus three actual). The exact reason for the discrepancy between our predictions and the actual TrES results could be one of many. As we discuss in the TrES section, from the properties of the known TrES planets, it may be that the survey has a brighter magnitude cut off than the one we use in our fiducial case. If this were the case, and we changed the magnitude limit from $m_R\leq13$ to $m_R\leq12$, the number of expected detections would drop to 8.16 from 13. Red noise also has the potential to substantially affect TrES; a specific estimate of the magnitude of red-noise in the TrES survey would be worthwhile.
 
Indeed, it is generally hard to draw specific statistical conclusions from the comparison of our results and those of the transit surveys because they typically do not adopt the strict detection criteria we have used. Promising planet candidates are often followed up even if they are beyond the stated S/N or magnitude limits of the survey. Understanding and quantifying how the survey teams select candidates is vital to appropriately deriving the statistical properties of extrasolar planets. Indeed, Tables 4-6 and 9-11 demonstrate that the actual predictions depend crucially on the specific magnitude and S/N threshold used.

As more transiting planets are discovered, these statistical properties are increasingly becoming the frontier of research, shifting the focus of the field away from individual detections. Given that transiting planets are so far one of the only ways that we may infer the radius of extrasolar planets and their exact orbital properties, the statistical characteristics of this group of objects is one of the only foreseeable ways that the areas of planetary interiors, system dynamics, migration, and formation will acquire more data. 

In the future, besides using the preceding formalism to design more efficient transit surveys, we would hope that a more systematic approach towards transit surveys will allow this model to be used to make more specific statistical comparisons. As the number of known transiting planets grows, we would also expect that this formalism will be used to test different distributions of planet frequencies, periods, and radii against those observational results. This will allow us to better understand the statistics of extrasolar planetary systems, improve our ability to find new planets, and help to understand the implications of the ones that have already been detected.

\acknowledgments 
We would like to thank the referee for a helpful report. We would also like to thank Dave Charbonneau and Francis O'Donovan for providing useful discussions and information about the TrES survey, Peter McCullough for his correspondence about the XO survey, and Scott Schnee for discussions about interstellar dust. TGB would like to thank Dave Latham and Josh Winn for helpful discussions and encouragement.

\appendix

\section{Photon Luminosity of a Source Star}
The photon luminosity in a bandpass of a given star of a given mass is calculated exactly by dividing the Planck radiation law for wavelength by the energy of a photon at that wavelength, multiplying the result by the surface area of the star, and then integrating over the wavelength range of the relevant bandpass:
\begin{equation}\label{eq:91}
{\Phi_{\lambda}(M) = \int_{\lambda_{min}}^{\lambda_{max}} \frac{B_\lambda (T_*)}{E_\lambda}} 4\pi^2 R_*^2 \, d\lambda, 
\end{equation}
Where $B_\lambda (T_*)$ is the Planck Law, and $E_\lambda = \frac{hc}{\lambda}$ is the energy of a photon. Note that this assumes that stellar radiation can be approximated as a blackbody, a crude assumption that still suffices for our purposes, as we show below.

Because the widths of the bandpasses in the visible and near-infrared that we are considering are generally narrow with respect to $\frac{dB_{\lambda}}{d\lambda}$, we can safely approximate this integral by the product of the central wavelength of each bandpass $\lambda_c$ and a bandpass width $\Delta \lambda$:
\begin{equation}\label{eq:101}
{\Phi_{\lambda}(M) = \frac{B_\lambda (T_*,\lambda_c) \Delta \lambda}{E_{\lambda_c}}} 4\pi^2 R_*^2 = \frac{8\pi^2 c R_*^2 \lambda_c^{-4} \Delta \lambda}{\mathrm{exp}[hc / \lambda_c k T_*(M)]-1},
\end{equation}
which, once we select a bandpass, only leaves the temperature of the source star $T_*(M)$ to be determined.

Using the Stefan-Boltzmann Law, we may relate the temperature of the source star to its bolometric luminosity and radius:
\begin{equation}\label{eq:111}
T_* = \left(\frac{L_{bol,*}}{4\pi \sigma R_*^2}\right)^{\frac{1}{4}},
\end{equation}
where $R_*$ and $L_{bol,*}$ are the radius and the bolometric luminosity of the source star, respectively. We chose to treat both as functions of the mass of the source star, as this is an independent variable that allows the use of the present day mass function to describe the general stellar population.

Using stellar data from \cite{harmanec1988} and \cite{popper1980}, we fit broken power laws for both the $R_*(M)$ and $L_{bol,*}(M)$ relations. For the mass-radius relation, we fit for two breaks in the power law, while the fitting for the mass-luminosity relation was accomplished using only one break. The relations that we thus found are:   

\begin{eqnarray}\label{eq:121}
R_*(M) = R_\odot \left(\frac{M}{M_\odot} \right)^\alpha\ \ \ \alpha&=&0.8\ \mbox{ for } M_* \leq 1,\\
\alpha&=&1.5\ \mbox{ for } 1 < M_* \leq 1.5, \nonumber \\
\alpha&=&.54\ \mbox{ for } 1.5 < M_*, \nonumber
\end{eqnarray}

\begin{eqnarray}\label{eq:131}
L_{bol,*}(M) = L_{bol,\odot} \left(\frac{M}{M_\odot}\right)^\beta\ \ \ \beta&=&2.5\ \mbox{ for } M_* \leq 0.4, \\
\beta&=&3.8\ \mbox{ for } 0.4 < M_*. \nonumber
\end{eqnarray}

Equations (\ref{eq:121}) and (\ref{eq:131}) are plotted against the \cite{harmanec1988} and \cite{popper1980} data in Figure 9.  

The real test, however, is how well these relations allow for the calculation of the effective temperature of a star at a given mass using (\ref{eq:111}), as this feeds directly into the determination of the photon luminosity in a given band. Figure 10 shows a plot of our mass vs. temperature relation along with data points from \cite{harmanec1988} and \cite{popper1980}. Additionally, we also show the analytical relation provided by \cite{harmanec1988}. One can see that our formulation of (\ref{eq:121}) and (\ref{eq:131}) affords a close approximation to both the data and the relation provided by \cite{harmanec1988}.

Substituting back into (\ref{eq:101}) allows us to calculate the photon luminosity of the source star, as well as the absolute magnitude of the source star in various bandpasses. If we compare these calculated absolute magnitudes in different bands for various stellar masses against empirical magnitudes for those same masses (Figures 11 and 12) we can see that the two relations are similar, to within half a magnitude, especially in longer wavelengths. Shorter wavelength bands such as U, B, and to a certain extent V, have calculated magnitudes that diverge from the empirical relation below about a solar mass, which is an unfortunate effect of using blackbody relations to determine the photon luminosity, as in these higher energy bands stellar spectra increasingly do not look like black bodies. Nevertheless, the blackbody assumption is sufficiently accurate for our purposes.

\section{Analytic Window Probability}

Roughly, if we ignore the effects of aliasing, we may say that the window probability of seeing a minimum of $N_{tr,min}$ transits of a planet with orbital period $p$ over the course of $N_n$ nights may be broken up into four different regimes.

First, if the average number of transits during the total time during which the target star is observed exceeds $N_{tr,min}$, then it is statistically ``certain'' that the transit will occur during the times of observation .

Second, it may be that the entire length of the survey is long enough that there are at least $N_{tr,min}$ transits, but that during the total time spent observing the average number of transits is less than that required for detection. Then one is not certain to detect $N_{tr,min}$ transits, and there is a decreasing window probability (for increasing planet period). 

Third, one may have $N_{tr,min}$ transits during an observing run, but only if the first transit occurs sufficiently early during that particular run. The probability of detecting the system then decreases at an even greater rate as the orbital period increases.

Finally, the orbital period of system may be so great that there is no possible way in which one could see the minimum number of transits required for a detection, and $P_{window}=0$. Mathematically then,
\begin{eqnarray}\label{eq:510}
P_{window} &=& 1\ \ \ \ \ \ \ \ \ \ \ \ \ \ \ \ \ \ \ \ \ \ \ \ \ \ \ \ \ \ \ \mbox{ for } 1 \leq \frac{<N_{tr}>}{N_{tr,min}},\\
&=& \left(\frac{<N_{tr}>}{N_{tr,min}}\right)^2\ \ \ \ \ \ \ \ \ \ \ \ \ \ \ \mbox{ for } \frac{<N_{tr}>}{N_{tr,min}} < 1 \leq \frac{(N_n-1)\Lambda+T_{n}}{N_{tr,min}P}, \nonumber \\
&=& \left(\frac{<N_{tr}>}{N_{tr,min}}\right)^2 \left(\frac{(N_n-1)\Lambda+T_{n}-(N_{tr,min}-1)P}{P}\right)^\frac{1}{2} \nonumber \\
&\ &\ \ \ \ \ \ \ \ \ \ \ \ \ \ \ \ \ \ \ \ \ \ \ \ \ \ \ \ \ \ \ \ \ \mbox{ for } \frac{(N_n-1)\Lambda+T_{n}}{N_{tr,min}P} < 1 \leq \frac{(N_n-1)\Lambda+T_{n}}{(N_{tr,min}-1)P}, \nonumber \\
&=& 0\ \ \ \ \ \ \ \ \ \ \ \ \ \ \ \ \ \ \ \ \ \ \ \ \ \ \ \ \ \ \ \mbox{ for } \frac{(N_n-1)\Lambda+T_{n}}{(N_{tr,min}-1)P} < 1, \nonumber
\end{eqnarray}
where $\Lambda = 1$ day. $<N_{tr}>$ is the average number of transits that will be seen during the time spent observing, and is given simply by:
\begin{equation}\label{eq:520}
<N_{tr}> = \frac{N_g T_n}{P}.
\end{equation}
$N_g$ is defined as the total number of ``good'' observing nights in a particular run, and may mathematically be described as $N_g=N_n-N_{lost}$, where $N_{lost}$ is the number of nights lost due to weather, technical problems, or other issues. $T_n$ is the length of each night's observing time. Alternatively, one may simply view the quantity $N_g T_n$ as the total amount of good observing time in a given run.

To test the accuracy of our formulation, we integrated and found the average window probability for both the analytic and exact formulations linearly averaged over several period ranges (1-3, 3-5, and 5-10 days). These are shown as a function of total observing nights under different viewing scenarios in Figure 13. 

One can see that as a result of the analytic solution not taking into account aliasing, there is divergence between the two solutions as the analytic form approaches unity, as well as in the cases requiring a minimum of three transits.

\section{The Present Day Mass Function}
A review of the literature shows that while there is agreement that the PDMF beyond $1M_\odot$ can be treated as a power law with a slope of $\alpha \approx5.3$, below a solar mass this agreement breaks down. This is because the low-mass end of the mass-function is unfortunately not as well constrained as it is for higher mass stars, reflecting the difficulty of observing dim M-dwarfs, as well as uncertainties in the mass-luminosity function \citep{henry1999}.   

For instance, \cite{reid2002}, who used the Hipparcos data sets along with their own data from the Palomar/Michigan State University (PMSU) survey to create a volume-limted sample out to 25 pc, propose a traditional power law formulation of the PDMF:
\begin{equation}\label{eq:350}
\Psi(M)=\frac{dn}{dM}= k_{PDMF} M^{-\alpha},
\end{equation}
with $\alpha = 5.2\pm0.4$ for $M>M_\odot$, and $\alpha = 1.35\pm0.2$ for $M\leq M_\odot$, and $k_{PDMF}$ as the correct normalization.

In a recent review, meanwhile, \cite{chabrier2003} argues for a log-normal function below a solar mass:
\begin{equation}\label{eq:360}
\Psi(M)=\frac{dn}{dM}= \frac{1}{M(\ln10)} A\ exp\left[\frac{-(\log M-\log M_c)^2}{2\sigma^2}\right],
\end{equation} 
with $A = 0.158_{-0.046}^{+0.051}$, $M_c = 0.079_{-0.016}^{+0.021}$, and $\sigma = 0.69_{-0.01}^{+0.05}$.

For comparison, if we normalize both to the \cite{scalo1986} normalization for 5 Gyr of $(dn/dM)_{1M_\odot} = k_{PDMF} = 0.019 M_\odot^{-1} \mathrm{pc}^{-3}$ and plot the two functions against each other (Figure 14), we can clearly see the divergence both numerically and functionally of these two forms below a solar mass. Interestingly however, for most surveys this divergence makes little difference in the final number of predicted detections: typically Reid's relation gives only 3\% fewer detections than Chabrier's does, if both are normalized at a solar mass. That this difference is so small comes from the fact that most wide-angle surveys are not sensitive to detections around stars with $\log(M)<-0.15$. Hence, the region of greatest uncertainty in the PDMF is left unseen by these surveys.

Since the difference between the two relations is small, and because the data set used by \cite{reid2002} is much more recent than that used by \cite{chabrier2003}, we therefore chose to use the \cite{reid2002} form of the PDMF.

For the purpose of comparing the Reid and Chabrier PDMFs in Figure 14, we normalized both to the value given by \cite{scalo1986}. Going forward, however, we change the normalization to the integrated mass density found observationally by \cite{reid2002} of $0.032 M_\odot \ \mathrm{pc}^{-3}$.

\section{Galactic Structure}
The values for the scale length and scale height of the Galactic disk that we use in our modeling are $h_{d,*} = 2.5\ \mathrm{kpc}$ and $h_{z,*}$ as a function of absolute magnitude. Specifically, we calculate the vertical scale height of the disk as:
\begin{eqnarray}\label{eq:370}
h_{z,*} &=& 90\ \ \ \ \ \ \ \ \ \ \ \ \ \ \ \ \ \ \ \ \ \ \ \ \ \mbox{ for } M_V \leq 2,\\
        &=& 90 + 200 \left(\frac{M_V-2}{3}\right)\ \mbox{ for } 2 < M_V < 5, \nonumber \\
        &=& 290\ \ \ \ \ \ \ \ \ \ \ \ \ \ \ \ \ \ \ \ \ \ \ \ \mbox{ for } 5 \leq M_V, \nonumber
\end{eqnarray}
which is a modified form of the relation found in \cite{bahcall1980}, who instead used a value of 320 pc for the scale height of stars with $5 \leq M_V$.

The large uncertainties in the value of the scale length of the disk reflects the uncertainty in the relevant literature. In their initial paper, \cite{bahcall1980} use $h_{d,*} = 2.5 - 3.0\ \mathrm{kpc}$. Since then various sources have proposed an even $h_{d,*} = 3.0\ \mathrm{kpc}$ \citep{kent1991}, have covered all the bases by using several values between $h_{d,*} = 2.0 - 3.0\ \mathrm{kpc}$ \citep{olling2001}, and have given the very exact result of $h_{d,*} = 2.264\pm.09\ \mathrm{kpc}$ \citep{drimmel2001}. 

To see what effect different scale lengths have on our final detection numbers, we simulated observations of a $6^\circ\times6^\circ$ field along the Galactic plane at intervals of $10^\circ$ in Galactic longitude, as well as fixing the Galactic longitude at certain points ($0^\circ$, $90^\circ$, and $180^\circ$) and varying the latitude of the fields. Figure 15 shows these results for scale lengths of 2, 2.5, and 3 kpc. At the extremes (directly towards or directly away from the Galactic core) the 2 and 3 kpc scale lengths differ by only 0.15 detections per field. We chose 2.5 kpc not only to split this difference, but also because more recent results place the scale length much lower than earlier 3 kpc estimates [e.g. - \cite{drimmel2001}].

In the case of the scale height of the Galactic disk, the literature divides between those who derive the scale height from optical star counts and those who rely on near-infrared observations. In the optical case, the scale height is presented as a function of absolute visual magnitude that varies from 90 pc to values between 250-320 pc. The near-infrared observations, on the other hand, lead to one single value of the scale height, generally around 290 pc. This happens because measurements in the infrared are dominated by older M-dwarfs that have dispersed in an even fashion away from the Galactic plane. 

To compare these different estimates, we have plotted the number of detections in a $6^\circ\times6^\circ$ field along $l=90^\circ$ as a function of Galactic latitude in Figure 16. We used two of the optically-derived scale height relations, as well as a constant scale height of 290 pc. One of the optical relations, the Bahcall model, is given by
\begin{eqnarray}\label{eq:375}
h_{z,*} &=& 90\ \ \ \ \ \ \ \ \ \ \ \ \ \ \ \ \ \ \ \ \ \ \ \ \ \mbox{ for } M_V \leq 2,\\
        &=& 90 + 230 \left(\frac{M_V-2}{3}\right)\ \mbox{ for } 2 < M_V < 5, \nonumber \\
        &=& 320\ \ \ \ \ \ \ \ \ \ \ \ \ \ \ \ \ \ \ \ \ \ \ \ \mbox{ for } 5 \leq M_V.\nonumber
\end{eqnarray}

The other optical model is taken from Allen's Astrophysical Quantities (AAQ) \citep{cox2000} and is: 
\begin{eqnarray}\label{eq:380}
h_{z,*} &=& 90\ \ \ \ \ \ \ \ \ \ \ \ \ \ \ \ \ \ \ \ \ \ \ \ \ \mbox{ for } M_V, \leq 2\\
        &=& 90 + 160 \left(\frac{M_V-2}{3}\right)\ \mbox{ for } 2 < M_V < 5, \nonumber \\
        &=& 250\ \ \ \ \ \ \ \ \ \ \ \ \ \ \ \ \ \ \ \ \ \ \ \ \mbox{ for } 5 \leq M_V. \nonumber
\end{eqnarray}

Also plotted in Figure 16 are the number of detections above and below $M_V=5$ for the three different scale height relations, as well as the residuals of the constant 290 pc and the AAQ relation against the Bahcall formulation. The exact choice of how the Galactic scale height is treated leads to rather different final predictions. Specifically, while the optical relations are similar, the constant 290 pc scale height from near-infrared observations leads to substantially more detections around stars with $M_V<5$. 

For our predictions, we have adopted the form of the \cite{bahcall1980} and Allen's Astrophysical Quantities scale height relations. We find the gradated scale height in these models appealing, because it accounts for the fact that more luminous stars with shorter lifetimes should not be dispersed far away from star-forming regions in the plane of the Galaxy. The constant 290 pc relation, because it is based on near-infrared data, reflects the sensitivity of those observations to less massive stars that are brighter in longer wavelengths. Nevertheless, we have used the 290 pc result as our value for the scale height of late-type stars with $5<M_V$, in contrast to the values used in the AAQ and Bahcall models.

\section{All-Sky Surveys' Limiting Magnitudes}

As per equation (\ref{eq:30}), we have
\begin{equation}\label{eq:440}
\sqrt{\chi^2} = S/N = (N_{tr})^{1/2}\ \frac{\delta}{\sigma}.
\end{equation}

For the all-sky surveys that we are considering, the number of in-transit data points observed is simply the number of data points acquired in each field multiplied by the fraction of time a planet will spend in transit during its orbit:
\begin{equation}\label{eq:450}
S/N = \left(\frac{t_{obs} \Theta_{FOV}}{t_{exp} \Omega_{survey}} \frac{R_*}{\pi a} \right)^{1/2} \frac{\delta}{\sigma},
\end{equation}
where $t_{obs}$ is the total observation time of the survey, $\Theta_{FOV}$ is solid angle of the survey telescope's field of view, and $\Omega_{survey}$ is the total solid angle of the sky covered by the survey.

Rearranging:
\begin{equation}\label{eq:460}
\sigma = \left(\frac{t_{obs} \Theta_{FOV}}{t_{exp} \Omega_{survey}} \frac{R_*}{\pi a} \right)^{1/2} \frac{\delta}{S/N}
\end{equation}

For Poisson, source-noise dominated errors, we have
\begin{equation}\label{eq:470}
\sigma = N_{source}^{-1/2}
\end{equation}
where $N_{source}$ is the total number of photons from the source collected during $t_{exp}$. We can then write,
\begin{equation}\label{eq:480}
\sigma = \left(\frac{t_{exp}\ e_\lambda}{t_{exp,0}}\right)^{-1/2} \left(\frac{D_{scope}}{D_{scope,0}}\right)^{-1} \sigma_0\ 10^{0.2(m_V-m_{V,0})}
\end{equation}
where $\sigma_0$ is the fiducial fractional photometric uncertainty achieved for a star with apparent magnitude $m_{V,0}$ using a telescope with diameter $D_{scope,0}$ and an exposure time of $t_{exp,0}$.  We have defined $e_\lambda$ as the efficiency of the observing setup relative to the fiducial case. If we rearrange equation (\ref{eq:480}) to
\begin{equation}
m_V = 5\log{\left[ \left(\frac{t_{exp}\ e_\lambda}{t_{exp,0}}\right)^{1/2}\left(\frac{D_{scope}}{D_{scope,0}}\right) \frac{\sigma}{\sigma_0} \right]} + m_{V,0}
\end{equation}
we may now substitute in equation (\ref{eq:470}) for $\sigma$. Doing so, and simplifying, we find that
\begin{equation}
m_V = 5\log{\left[ \left(\frac{t_{obs}\ e_\lambda}{t_{exp,0}} \frac{\Theta_{FOV}}{\Omega_{survey}} \frac{R_*}{\pi a}  \right)^{1/2} \frac{D_{scope}\ \delta}{D_{scope,0}\ \sigma_0\ S/N} \right]} + m_{V,0}
\end{equation}
Which is the limiting magnitude around which an all-sky survey will detect planets.

\clearpage

\begin{figure}
\vskip -0.1in
\epsscale{0.75} 
\plotone{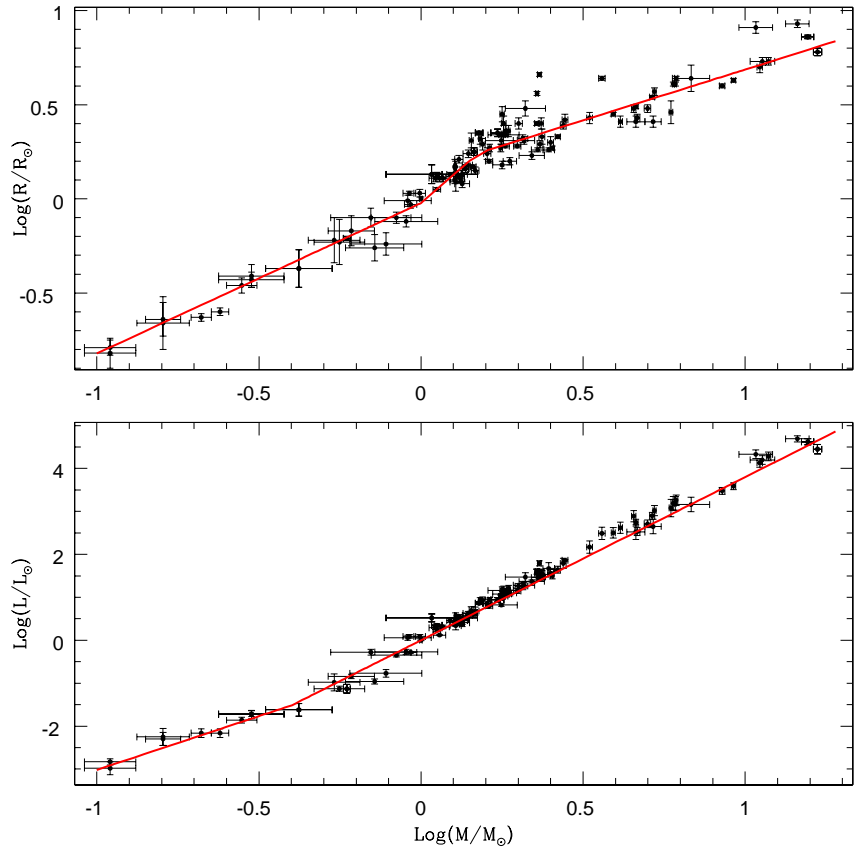}
\vskip -0.0in 
\figcaption[]{The mass-radius and mass-luminosity relations. The data points from \cite{harmanec1988} and \cite{popper1980} are overlaid with our best fit broken power laws for each. Note the the mass-radius fit has two breaks as opposed to only one in the mass-luminosity fit.}
\end{figure}

\begin{figure}
\vskip -0.25in
\epsscale{0.75} 
\plotone{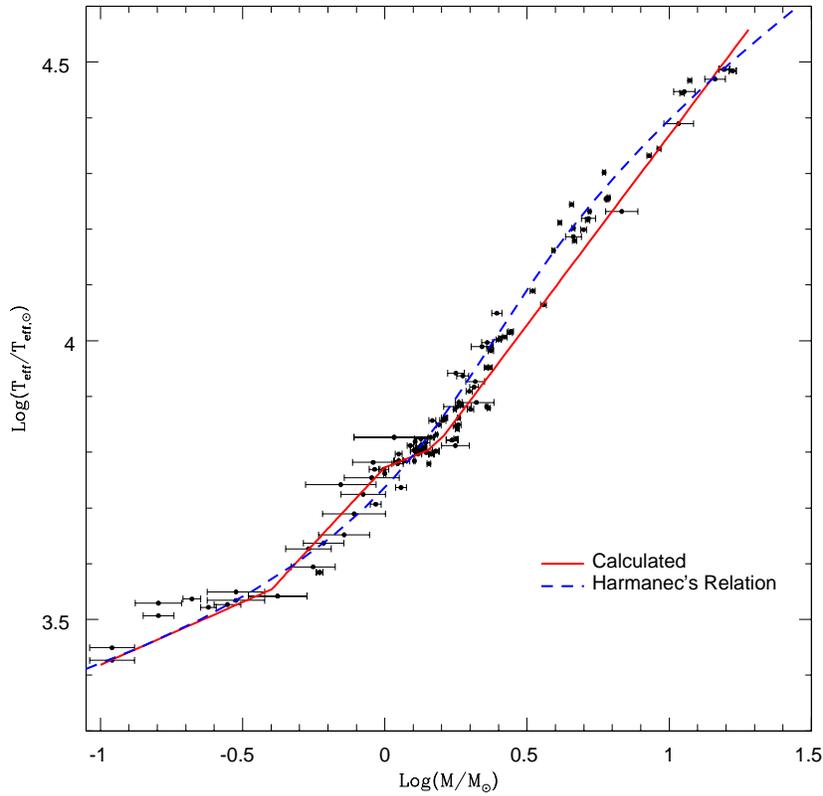}
\vskip -0.1in 
\figcaption[]{The mass-temperature relation, again using data from \cite{harmanec1988} and \cite{popper1980}. Overlaid onto the plot are our calculated relationship, as well as the analytical relation described by \cite{harmanec1988}.}
\end{figure}

\begin{figure}
\vskip -0.15in
\epsscale{0.75} 
\plotone{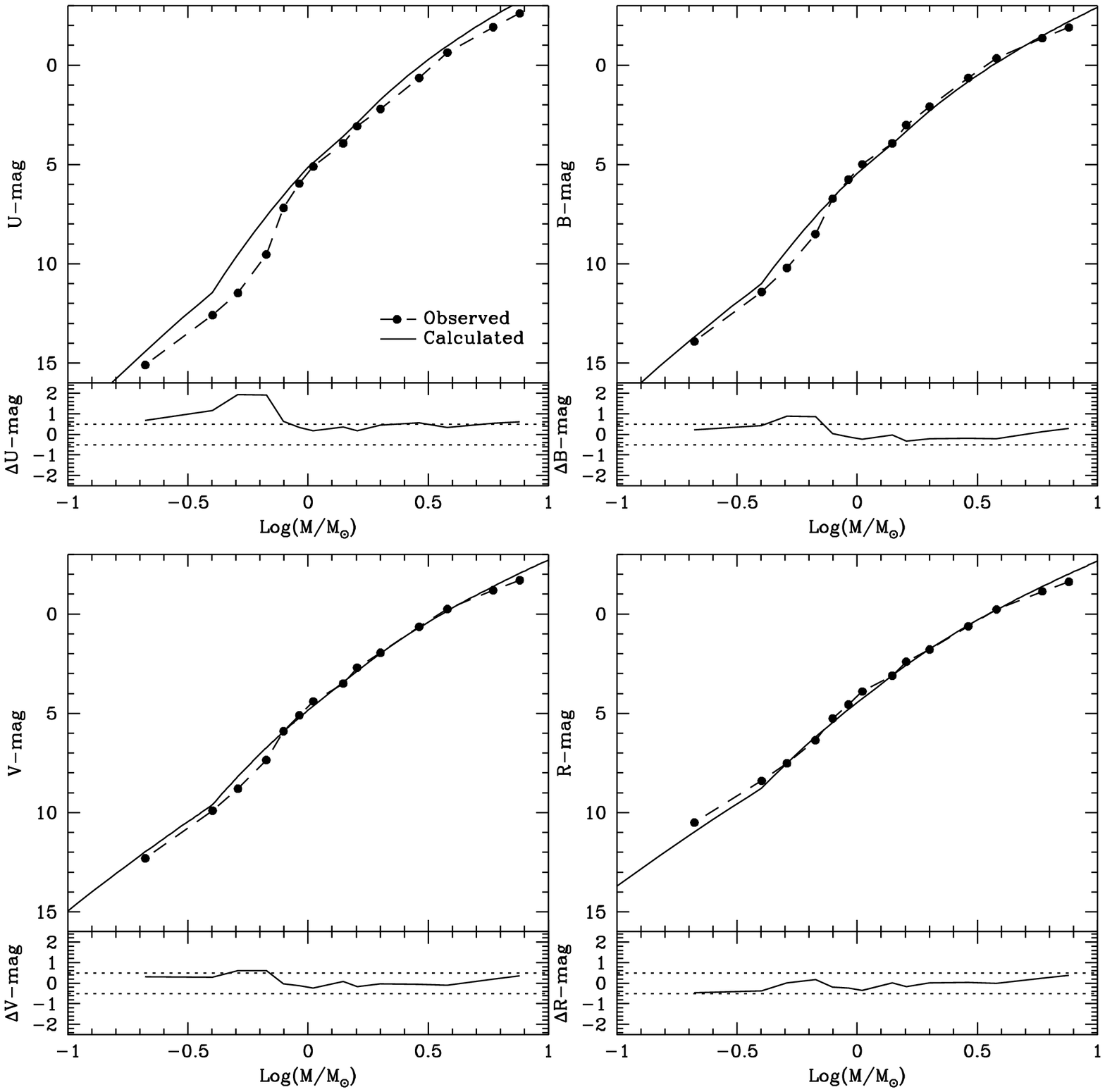}
\vskip -0.0in 
\figcaption[]{Our calculated U-, B-, V-, and R-band magnitudes as a function of mass, compared to observed stellar magnitudes in those bands \citep{cox2000}. Below each is a plot of the residuals in that band.}
\end{figure}

\begin{figure}
\vskip -0.2in
\epsscale{0.75} 
\plotone{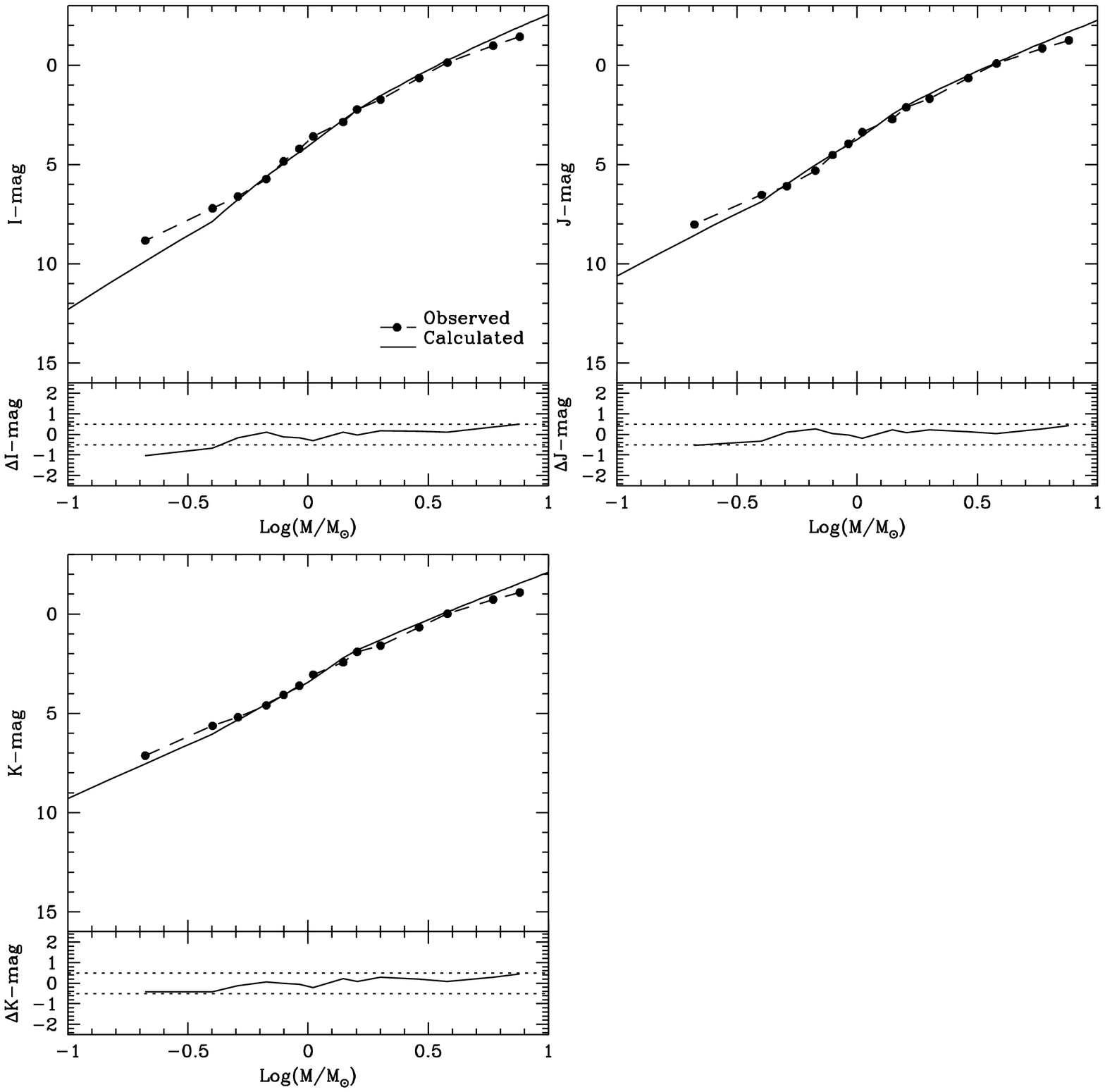}
\vskip -0.0in 
\figcaption[]{Our calculated I-, J-, and K-band magnitudes as of function of mass, compared to observed stellar magnitudes in those bands \citep{cox2000}. Below each is a plot of the residuals in that band.}
\end{figure}

\begin{figure}
\vskip -0.15in
\epsscale{0.75} 
\plotone{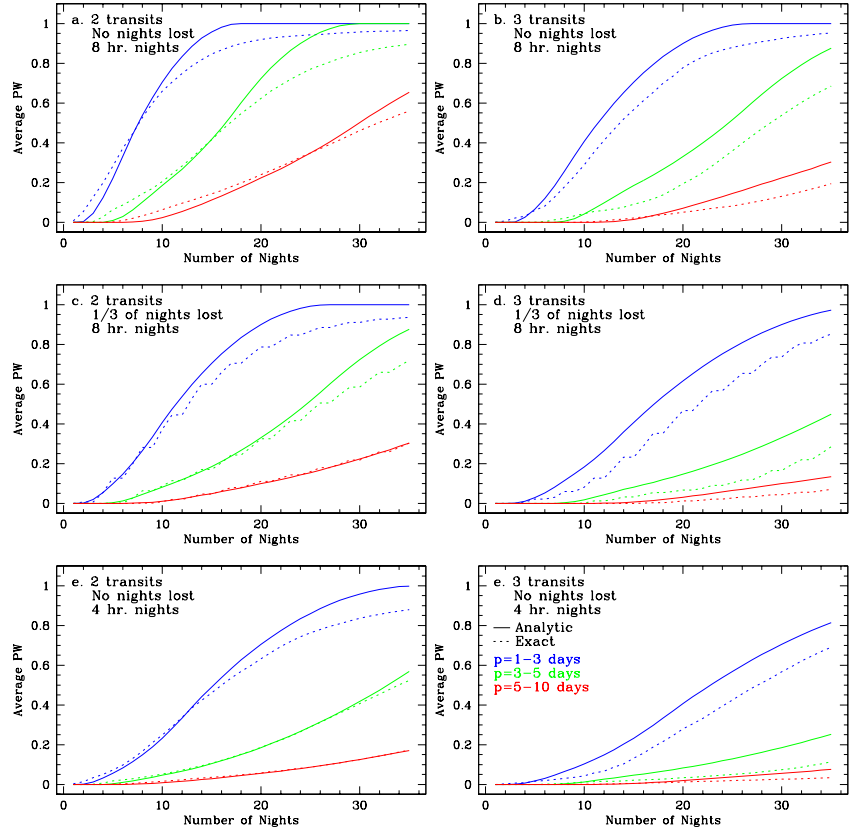}
\vskip -0.1in 
\figcaption[]{A comparison of our analytic formulation of the window probability versus using an exact formulation to calculate the average window probability as a function of total observing nights averaged over three different orbital period ranges assuming a linear distribution of periods. The left side uses a two transit minimum, while the right requires three transits. The top two plots - a) and b) - use the idealized conditions of 8 hour nights and no observing losses due to weather. The middle plots - c) and d) - also use 8 hour nights, but now 1/3 of the total number of nights are lost due to bad weather. Finally, the bottom two plots - e) and f) - show the effects of shorter 4 hour nights but no losses due to weather.}
\end{figure}

\begin{figure}
\vskip -0.3in
\epsscale{0.75} 
\plotone{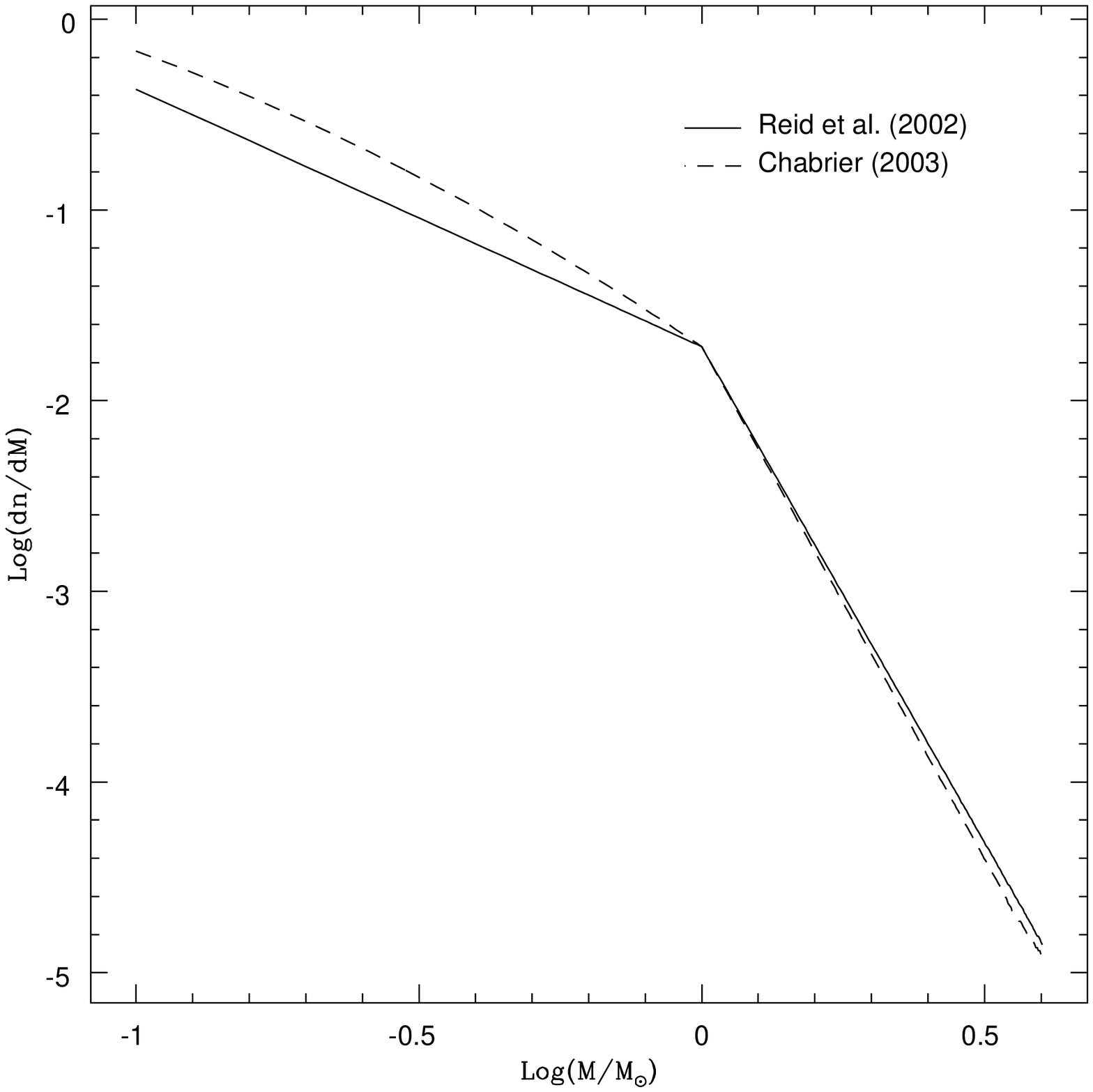}
\vskip -0.1in 
\figcaption[]{A comparison of the PDMFs proposed by \cite{reid2002} and \cite{chabrier2003}. \cite{reid2002} uses a power law (equation [\ref{eq:350}]) below a solar mass, while \cite{chabrier2003} uses a log-normal function (equation [\ref{eq:360}]).}
\end{figure}

\begin{figure}
\vskip -0.2in
\epsscale{0.73} 
\plotone{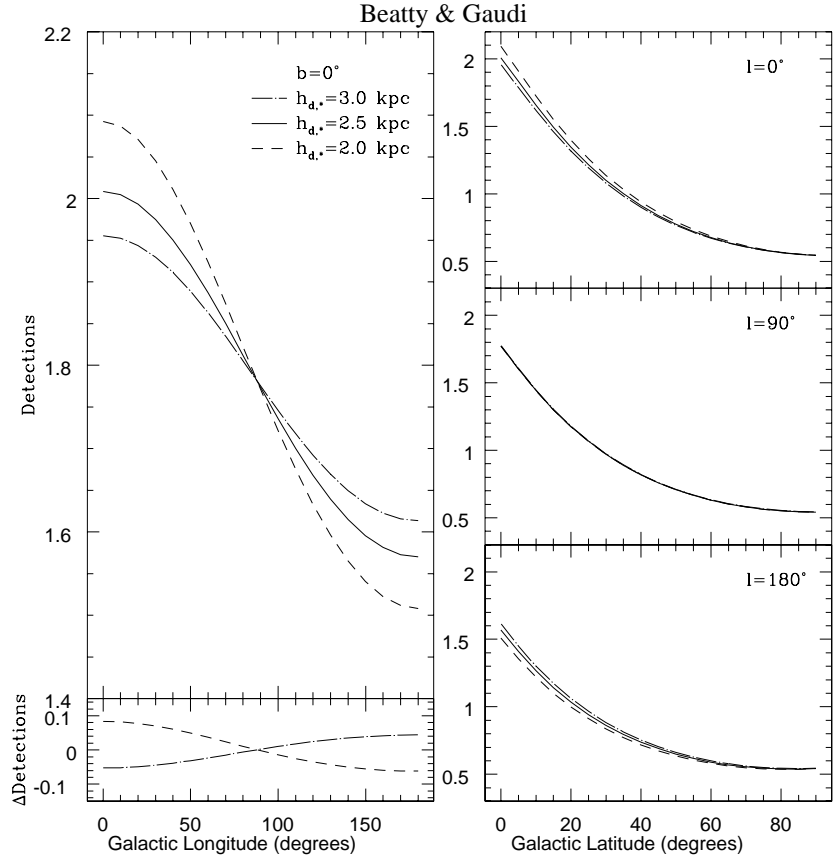}
\vskip -0.0in 
\figcaption[]{The effects of various scale lengths on the predicted number of detections in a $6^\circ\times6^\circ$ field. The left shows detections as a function of Galactic longitude along the Galactic plane, as well as the difference in the number of detections from the assumption of a scale height of 2.5 kpc. The right plots detections as a function of Galactic latitude for three different Galactic longitudes. Note the lack of effect of the scale length at $l=90^\circ$, as well as the convergence of all three at $b=90^\circ$, as is to be expected.}
\end{figure}

\begin{figure}
\vskip -0.2in
\epsscale{0.73} 
\plotone{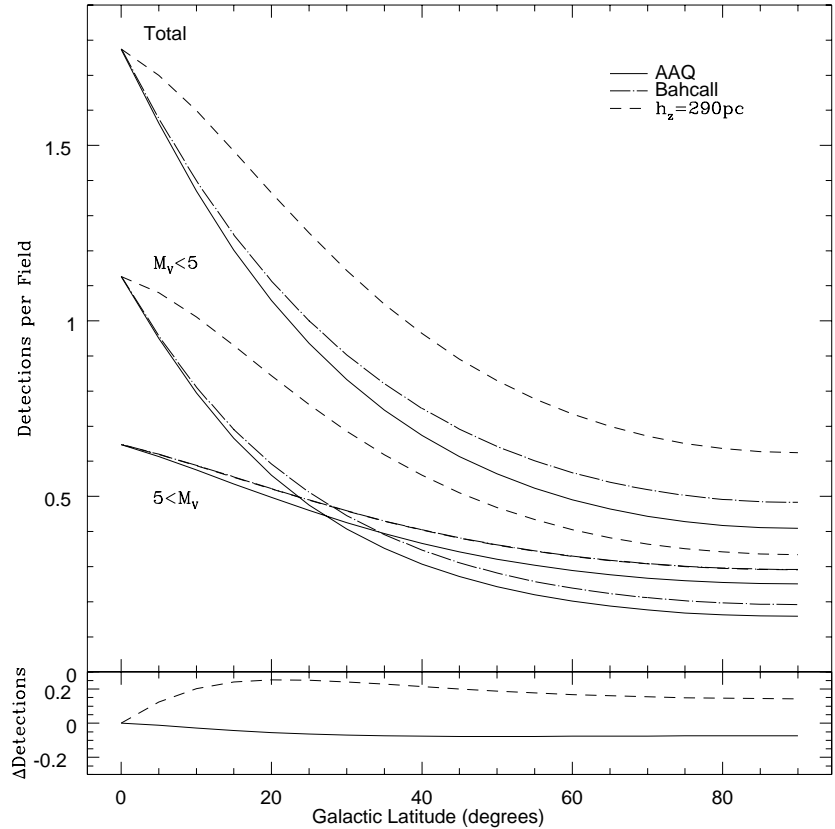}
\vskip -0.0in 
\figcaption[]{The total number of detections in a $6^\circ\times6^\circ$ field as a function of Galactic latitude for a variety of scale height relations. Also shown are the number of detections around stars with $M_V<5$ and $5<M_V$, as well as the difference in the predicted number of detections using the constant 290 pc and Allen Astrophysical Quantities \citep{cox2000} scale height relations relative to the predictions using \cite{bahcall1980}.}
\end{figure}

\clearpage

\end{document}